\documentclass[usenatbib]{mn2e}
\input{psfig}
\newdimen\hssize
\newdimen\hdsize
\newcommand{\beq}{\begin{eqnarray}}
\newcommand{\eeq}{\end{eqnarray}}
\newcommand{\zz}{$z\sim 3$}

\newcommand{\gal}{{\sc galics}}
\newcommand{\momaf}{{\sc momaf}}

\newcommand{\mo}{{\sc momaf}}
\newcommand{\hatton}{{\sc galics~i}}
\newcommand{\devriendta}{{\sc galics~ii}}
\newcommand{\devriendtb}{{\sc galics~iv}}

\newcommand{\blaizotb}{{\sc galics~v}}

\newcommand{\apj}{ApJ}

\newcommand{\apjl}{ApJL}
\newcommand{\aj}{AJ}
\newcommand{\mnras}{MNRAS}
\newcommand{\aap}{A\&A}

\newcommand{\araa}{ARA\&A}

\begin{document}

\title[GALICS III : Predicted Properties for LBGs at $z\sim 3$]
      {GALICS III : Predicted Properties for Lyman Break Galaxies at redshift 3}
\author[Blaizot et al.]
       {J\'er\'emy Blaizot$^{1,2}$, Bruno Guiderdoni$^1$, Julien E. G.
Devriendt$^{2,1}$, \and
Fran\c{c}ois R. Bouchet$^1$, Steve J. Hatton$^1$, and Felix Stoehr$^3$\\
$^1$ Institut d'Astrophysique de Paris, 98 bis boulevard Arago, 75014 Paris, 
France. \\
$^2$ Oxford University, NAPL, Keble Road, Oxford OX1 3RH, United Kingdom.\\
$^3$ Max-Planck-Institut f\"ur Astrophysik, Karl-Schwarzschild-Str. 
1, 85741 Garching, Germany.}	

\date{}
\pagerange{\pageref{firstpage}--\pageref{lastpage}}
\pubyear{2003}
\maketitle
\label{firstpage}

\begin{abstract}
This paper illustrates how mock observational samples of
high--redshift galaxies with sophisticated selection criteria can be
extracted from the predictions of \gal, a hybrid model of hierarchical
galaxy formation that couples the outputs of large cosmological
simulations, and semi--analytic recipes to describe dark matter
collapse, and the physics of baryons respectively. As an example of
this method, we focus on the properties of Lyman Break Galaxies at
redshift $z \sim 3$ (hereafter LBGs) in a $\Lambda$CDM cosmology.
With the \mo~software package described in a companion paper, we
generate a mock observational sample with selection criteria as
similar as possible to those implied in the actual observations of $z
\sim 3$ LBGs by \citet{SteidelPettiniHamilton95}. We need to introduce
an additional ``maturity'' criterion to circonvene subtle effects due
to mass resolution in the simulation. We predict a number density of
1.15 arcmin$^{-2}$ at $R \leq 25.5$, in good agreement with the
observed number density 1.2 $\pm 0.18$ arcmin$^{-2}$. Our model allows
us to study the efficiency of the selection criterion to capture $z
\sim 3$ galaxies. We find that the colour contours designed from
models of spectrophotometric evolution of stellar populations are able
to select more ``realistic'' galaxies issued from models of
hierarchical galaxy formation. We quantify the fraction of interlopers
(12 \%), and the selection efficiency (85\%), and we give estimates of
the cosmic variance.  We then study the clustering properties of our
model LBGs.  They are hosted by halos with masses $\sim 1.6 \times
10^{12} M_\odot$, with a linear bias parameter that decreases with
increassing scale from $b=5$ to $3$. The amplitude and slope of the 2D
correlation function is in good agreement with the data. We
investigate a series of physical properties: UV extinction (a typical
factor 6.2 at 1600 \AA{}), stellar masses, metallicities, and Star
Formation Rates, and we find them to be in general agreement with
observed values.  The model also allows us to make predictions at
other optical and IR/submm wavelengths, that are easily accessible
though queries to a web interfaced relational database.  Looking into
the future of these LBGs, we predict that 75 \% of them end up as
massive ellipticals and lenticulars today, even though only 35 \% of
all our local ellipticals and lenticulars are predicted to have a LBG
progenitor. In spite of some shortcomings that come from our
simplifying assumptions and the subtle propagation of mass resolution
effects, this new 'mock observation' method clearly represents a 
first step toward a more accurate comparison between
hierarchical models of galaxy formation and real observational
surveys.
\end{abstract}

\section{Introduction} \label{sec:intro}
Models of hierarchical galaxy formation are now sophisticated enough
to produce a host of predictions for the statistical properties of
local and high--redshift galaxies. These models have to be tested
against observational samples. However, such a comparison is not as
straightforward as it might appear at first glance, because
observations are affected by selection criteria and observational
biases. The difficulty of the comparison is enhanced for samples of
high--redshift galaxies that are selected only on the basis of their
apparent magnitudes and colours, that is, according to properties that
are far from the dynamical quantities models of galaxy formation deal
with. In such a case, it is not easy to track back the propagation of
the selection criteria and observational biases to the ``model''
parameter space. At this stage, the best method consists in putting
model predictions into the ``observation'' parameter space, by
producing mock galaxy samples that are obtained with selection
criteria and observational biases as close as possible to those that
affect the real observational samples.

This paper uses the predictions of the \gal~model of hierarchical
galaxy formation (for {\em Galaxies in Cosmological Simulations}) that
embodies the so--called ``hybrid approach'' by coupling the
description of dark matter collapse by means of cosmological $N$--body
simulations, and the description of the physics of baryons with
semi--analytic recipes, including the estimate of spectral energy
distributions in the UV to sub-millimetre wavelength range
(\citet{HattonEtal03}, hereafter \hatton, and \citet{DevriendtEtal03},
hereafter \devriendta). We use the \mo~package (for {\em Mock Map
Facility}) described in \citet{BlaizotEtal03a} (hereafter \momaf) to
produce mock observing cones from which mock samples can be extracted,
with selection criteria and biases that mimic those of actual
observations.

We hereafter illustrate such a method by addressing the constraints put
on models by the so--called {\em Lyman Break Galaxies} (hereafter
LBGs) at redshift $z \sim 3$ \citep{SteidelEtal96}.  These galaxies
are obtained using the UV drop--out technique, first employed for the selection of distant
galaxies by \citet{SteidelHamilton93}, that simply relies on
the shift of the Lyman break through broad--band filters. This break,
being mainly due to the absorption of the UV photons by the H{\sc i}\
of the intergalactic medium (IGM), is roughly independent from the
galaxies' intrinsic properties, thus allowing the efficient selection
of a complete sample of luminous galaxies at a given redshift, with
photometry in only three broad--band filters. In the last years, this
technique has led to a tremendous increase in the number of observed
objects at $z\sim 3$.  Although the set of data is impressively large
today, compared to only a few years ago, a theoretical picture has not
yet clearly emerged. This is mainly due to a poor understanding
of the interplay of star formation, feedback and extinction on
cosmological scales within hierarchical clustering, and also to the
difficulty to compare ``unbiased'' and ``noiseless'' theoretical work
with ``corrected--for--everything'' data. In spite of these
difficulties, the predictions of the properties of LBGs obtained from
semi--analytic models of hierarchical galaxy formation
(\citet{BaughEtal98}, \citet{SomervillePrimackFaber01}, \citet{IdziEtal03}) or
fully numerical approaches (\citet{Nagamine02}, \citet{WeinbergHernquistKatz02}) are
encouraging. 

We hereafter produce a mock sample of $z \sim 3$ LBGs from our model
of hierarchical galaxy formation, by applying selection criteria and
observational biases that are as similar as possible to those involved
in the definition of the data gathered by \citet{SteidelEtal96}. The
comparison of our mock sample with the actual sample has a three--fold
interest. {\it First}, it helps us to explore the selection criteria with a
plausible model of galaxy formation, whereas the UV--drop out
technique has been primarily designed within the paradigm of
monolithic galaxy formation.  This allows us to get some insight on
the efficiency of the selection criteria to capture {\em bona fide}
galaxies at $z \sim 3$, as well as to assess the number of
interlopers, that is, of galaxies that pass through the criteria, but
are not at the target redshift $z \sim 3$. In addition we study 
the field--to--field
cosmic variance.  {\it Second}, we compare predictions of our model 
of hierarchical galaxy formation with
observational properties. Since \gal~has specific features such as
spatial information, multi--wavelength predictions from the
(rest-frame) UV to sub-mm, and hierarchical information on the merging
history trees, we can address questions such as clustering,
optical/IR luminosity budget, and current descendants of $z \sim 3$
LBGs. {\it Third}, our predictions can be used as an educated 
guideline for extensive follow--up observations of LBGs at other optical 
and IR/submm wavelengths. The overall
agreement of the predictions with the data, in spite of obvious
shortcomings, encourages us to think that the picture of hierarchical
galaxy formation implemented in our \gal~model gives a good
description of optically--bright, star forming
galaxies at $z \sim 3$. We note however that not all galaxies at this
redshift are correctly reproduced by \gal.

This paper is the third in the \gal~series after paper I (\hatton) that
presented the global features of our model and a set of properties for
local galaxies, and paper II (\devriendta) that showed predictions for
hierarchical galaxy evolution between redshifts 0 and 3. It is also an
illustration of the generic method of mock map making detailed in
\momaf.  Predictions of faint galaxy counts and 2D clustering for
the global population will be given in forthcoming paper
IV (Devriendt et al. 2003b, hereafter \devriendtb), and paper V
(Blaizot et al. 2003b, hereafter \blaizotb) respectively. These five 
\gal~papers,
together with the \mo~paper, complete the presentation of the first version of
the \gal~project. Results of the model, as well as extensive follow--up 
predictions at many optical and IR/submm wavelengths, are publicly 
available from a web interfaced relational database
located at the URL {\tt http://galics.iap.fr}.

This paper is organised as follows. Section 2 gives a brief summary of
the features of \gal~and \mo~that are relevant for this study. The
selection criteria are applied to our mock observing cone in section
3, where we explore various aspects of the UV--drop out
technique. Section 4 gives predictions for the cosmic variance, and
compares the predicted clustering with observational data.  The 
UV to IR luminosity
budget is studied in section 5, where we try to clarify the
controversial issue of extinction in LBGs from a theoretical point of
view. An attempt to elucidate the nature of LBGs, and their fate at
the present time within the context of hierarchical clustering, is
given in section 6. Finally, section 7 gives a summary of the results, 
and discusses
their robustness in light of the shortcomings of the method.

\section{Simulating observations} \label{sec:SimObs}

\subsection{A brief summary of the \gal{} model}
\gal{} is a model of
hierarchical galaxy formation which combines high-resolution
cosmological simulations to describe the dark matter content of the
Universe with semi-analytic prescriptions to deal with the baryonic
matter. This hybrid approach is fully described in \hatton{} and \devriendta{},
so we only briefly recall its main characteristics hereafter.

\subsubsection{Dark matter simulation} \label{sec:dm_sim}
The cosmological N-body simulation we refer to throughout this paper
was carried out using the parallel tree-code developed by \citet{Ninin99}. We use
 a flat Cold Dark Matter model with a cosmological constant
($\Omega_m = 0.333$, $\Omega_{\Lambda} = 0.667$). The simulated volume
is a cube of side $L_{box}=100h_{100}^{-1}$Mpc, with $h_{100}^{-1} =
0.667$, containing $256^3$ particles of mass $8.3\times 10^9$M$_{\odot}$,
with a smoothing length of 29.29 kpc. The power spectrum was set in
agreement with the present day abundance of rich clusters ($\sigma_8 =
0.88$), and we followed the DM density field from $z=35.59$ to $z=0$,
outputting 100 snapshots spaced logarithmically in the expansion factor.

On each snapshot we use a friend-of-friend algorithm to identify
virialised groups of more than 20 particles, thus setting the minimum
dark matter halo mass to $1.66\times 10^{11}$ M$_{\odot}$. We compute a set
of properties of these haloes, including position and velocity of the
centre of mass, kinetic and potential energies, and spin
parameter. Assuming a singular isothermal sphere density profile for our
virialised dark matter haloes, we then compute their virial radius, enforcing
conservation of mass and energy.

Once all the haloes are identified in each snapshot, we compute their
merging history trees, following the constituent particles from one
output to the next. The merging histories we obtain are by far
more complex than in semi-analytic approaches as they include
evaporation and fragmentation of haloes. The way we deal with these is
described in details in \hatton.

\subsubsection{Baryonic prescriptions} \label{sec:baryonic prescriptions}
When a halo is first identified, it is assigned a mass of hot gas,
assuming a universal baryon to dark matter mass ratio ($\Omega_B/\Omega_0 = 0.135$ in our
fiducial model). This hot gas is assumed to be shock
heated to the virial temperature of the halo, and in hydrostatic
equilibrium within the dark matter potential well. The comparison of
the cooling time of this gas to its free-fall time, as a function of
the radius, yields the mass of gas that can cool to a central disc
during a time-step. The size of this exponential disc is given by
conservation of specific angular momentum during the gas infall and
scales linearly with the spin parameter of the halo. Cold
gas is then transformed into stars at a rate
inversely proportional to the dynamical timescale, with a constant
efficiency $\beta^{-1}$. Newly formed stars are distributed
according to the \citet{Kennicutt83} initial mass function (IMF).
These stars are then evolved between 
timesteps, using a sub-stepping of at most 1 Myr. During each sub-step,
a fraction of the stars explode as supernovae, releasing metals into the ISM.
Energy is also released, which heats the ISM and 
may blow part of it away into the IGM, with some efficiency $\epsilon$.

When two haloes merge, the galaxies they contain are placed in the
descendant halo. Their initial radial positions in this new halo
are obtained through a perturbation of the final radial positions they had
in each of the progenitor haloes. Due to subsequent dynamical friction or 
satellite-satellite encounters, galaxies can then merge in their new host 
halo. When such events occur, a 'new' galaxy is then created (which is    
the descendant of the two progenitor galaxies that have merged) and       
the stellar and gaseous contents of its three components (disc, bulge     
and starburst) are deduced from those of its two progenitors using a      
single free parameter $\chi$. Note that the descendant galaxy can
either become elliptical (in shape) or retain a large disc depending
on the mass ratio of its progenitors. After a merging, provided gas is
available, a nuclear starburst is assumed to take place at the
centre of the 'new' galaxy. The radius of this burst is fixed to be
$\kappa$ times the radius of the 'new' bulge.

The spectral energy distributions (SEDs) of our modelled galaxies are
computed by summing the contributions of all the stars they contain,
according to their age and metallicity, both of which we keep track
of in the simulation. Extinction is computed assuming a
random inclination for disc components, and the emission of dust is
added to the extinguished stellar spectra 
in a similar way as prescribed in the {\sc stardust} code
\citep{DevriendtGuiderdoniSadat99}. Finally, a mean intergalactic medium (IGM)   
extinction correction is implemented following the procedure described in
\citet{Madau95}, before we convolve the SEDs with the desired observer
frame filters (see \momaf{}).

Thus our model depends only on a few free parameters, which are given the 
standard values quoted in \hatton{} and \devriendta{}, 
namely, $\beta^{-1}=0.02$, $\epsilon = 0.1$, 
$\chi = 3.333$, $\kappa = 0.1$. As it is explained in \hatton{}, the other 
two free parameters (the normalisation $\psi$ of the frequency of 
satellite--satellite encounters, and the efficiency $\zeta$ for the 
recycling of metals ejected into the intergalactic ``reservoir'') 
only have a minor role. 

\subsection{Resolution effects} \label{sec:res_effect}
The mass resolution of the N-body DM simulation we use has important
repercussions on the statistical and physical properties of our
modelled galaxies. We outline the main two effects below.

\subsubsection{Magnitude completeness limit}
The mass resolution of the N-body simulation is, for our concerns, the
minimum mass of a halo, that is, the mass of 20 particles ($\sim 1.6\times
10^{11} M_{\odot}$).  It is the mass of the smallest
structure in which we may form a galaxy. This halo mass corresponds to 
a galaxy mass $M_{res}=2.2\times 10^{10} M_{\odot}$ assuming that all the gas in
the halo can cool. Galaxies with a mass higher than this cannot exist in
unresolved haloes, and our sample of such galaxies is thus complete.
On the contrary, galaxies with a lower mass can (and do) exist, but
our sample is not complete since we miss all those that would lie in
unresolved haloes.

\begin{figure}
\centerline{\psfig{figure=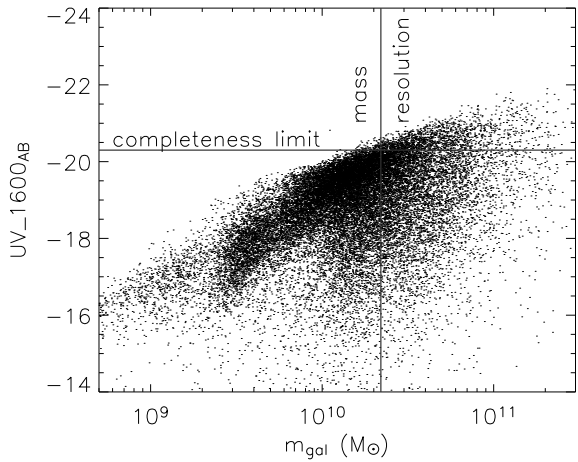,width=\hssize}}
\caption{Absolute AB magnitudes of $z=3$ galaxies at rest-frame 1600 \AA\ as a
function of baryonic mass. The
absolute magnitudes take into account dust absorption. The
vertical line indicates our mass resolution, $M_{res}=2.2\
10^{10}M_{\odot}$, above which our sample of galaxies is complete. 
The horizontal line represents our completeness
limit in terms of magnitude so that
galaxies brighter than $M_{AB}^{res}(1600) = -20.3$ also form a
complete sample. }
\label{fig:res_effect}
\end{figure}

It is useful to convert this completeness limit in terms of magnitudes
in order to understand in which range of luminosity our predictions
are reliable (e.g. for the luminosity functions in section
\ref{sec:luminosity_functions}). To do this, we plot the absolute
rest-frame UV magnitude at 1600\AA\ (internal dust absorption included)
versus the total baryonic mass of galaxies in Fig.
\ref{fig:res_effect}, directly from the $z=3$ simulation snapshot.
Note that the large scatter on this plot is due to the fact
that at this wavelength, we see the combined effects of most recent star
formation and dust extinction, which are poorly correlated with the total 
mass of stars.  We define the magnitude completeness limit $M_{AB}^{res}(1600)$ so that 90\% of
the galaxies brighter than $M_{AB}^{res}(1600)$ have a mass greater
than $M_{res}$.  As shown in Fig. \ref{fig:res_effect}, we expect
our sample of galaxies brighter than $M_{AB}^{res}(1600) = -20.3$ to
be complete at $z\sim 3$.

The magnitude completeness limit of our simulation is thus comparable
to the detection limit for Steidel's sample of LBGs (absolute
magnitude $R \leq -20.2$ in our $\Lambda$CDM cosmology), which allows us to
compare our results with their data.

\subsubsection{History resolution} \label{sec:hist_resolution}
A more subtle effect of resolution is that missing small structures
means missing part of our halo merging histories. As a matter of fact,
the impact on our modelled galaxies is twofold :
\begin{itemize}
\item we miss small galaxy mergers because galaxies  
  close to the resolution limit are formed smoothly in  
  haloes at the resolution limit, instead of assembled through mergers of smaller objects
  that would have populated smaller progenitor haloes;
\item we cool too much gas in too short a time on new central galaxies in
  haloes at the resolution limit. Had we resolved their host halo merging
  histories, part of this cold gas would have been split between smaller progenitors 
  at earlier times and partly processed into stars already.
\end{itemize}
However, the 'observational' signature of this resolution effect (physical properties, 
luminosities of galaxies) tends to disappear after a while.  In practice,
a galaxy in our standard N-body simulation needs to evolve for
about 1 Gyr before we can be sure that its properties have converged. 
This is irrelevant at $z=0$ (resp. $z=1$), where only 3 (resp. 69) out
of 19,372 (resp. 19,257) galaxies above the formal resolution limit
$M_B=-18.9$ (resp. $M_B=-20.2$) are younger than this threshold. However, it
becomes problematic at $z=3$, when the age of the Universe is
only a couple of gigayears old and roughly 75\% of the galaxies are 
concerned.  As a
consequence, we introduce in the following study an additional free parameter
that we call the {\it maturity age} $t_{mat}$, and we 
discard from our sample any LBG whose first progenitor is younger than
$t_{mat}=1.1$ Gyr at the time when this LBG is identified as such. As the need 
for this kind of selection is one of the most important drawbacks of our study,
section~3.2.2 justifies in detail our choice of maturity criterion.

\subsection{A typical mock field}
Because Lyman Break selection is by definition an observational
selection, it is necessary to apply as similar as possible criteria
to our modelled galaxies sample before attempting a thorough comparison with
the data. We therefore used the \momaf~package to simulate light-cones
from \gal~outputs, as described in the \momaf{} paper (see also the
\gal/\momaf~web-page {\tt http://galics.iap.fr/}).  Basically, the
method consists in tiling a time sequence of simulation boxes along a mock line of sight,
and computing the apparent properties of galaxies from their
positions in this cone-like geometry. It allows us to use as much of the
spatial information contained in the N-body simulation as possible in
order to produce realistic mock catalogues, which can then be analysed
in the same way as real observations.

\begin{figure}
\centerline{\psfig{figure=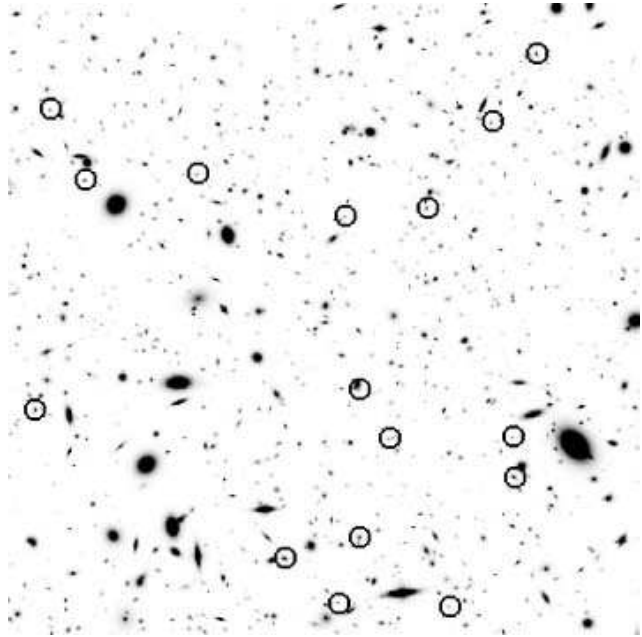,width=\hssize}}
\caption{Mock medium deep field, of size $3 \times 3$ arcmin$^2$, seen in the $R$ band. 
Circles mark the position of LBGs which were selected using 
the colour and apparent magnitude criteria given by Eq. \ref{eq:Steidel_contour}.}
\label{fig:mock_hdf}
\end{figure}

In Fig. \ref{fig:mock_hdf}, we show a typical result of this work :
a mock medium deep field of $3\times 3$ arcmin$^2$ seen in the $R$ band
\citep{SteidelHamilton93}. The circles indicate LBGs, which are identified as
indicated in section \ref{sec:ccplot}. 

\section{LBG selection process}

\subsection{Colour-colour selection} \label{sec:ccplot}
It is the IGM extinction that makes the Lyman break selection such an
efficient technique for selecting high-redshift galaxies. Galaxy
evolution, along with dust absorption inside galaxies, also produce a
Lyman-break, but its intensity varies from one galaxy to
the next, so that galaxies end up scattered all over the colour-colour plane 
and as a result, colour and redshift would be strongly degenerate.
Not only does the IGM attenuation yield a more pronounced Lyman-break, but the fact that it
is due to material {\it external} to the galaxies makes this break
independent from intrinsic properties of the observed objects, allowing
the observer to select a whole population of ``normal'' galaxies at \zz.

\begin{figure}
\centerline{\psfig{figure=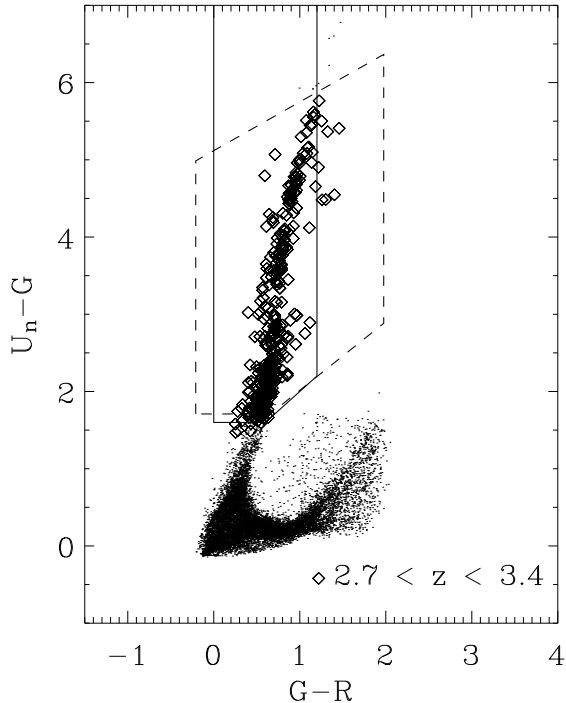,width=\hssize}}
\caption{$(U_n - G)$-$(G-R)$ diagram, for a mock 360 arcmin$^2$ field. 
Only galaxies with apparent magnitude $R_{AB} \leq 25.5$ are shown.
Diamonds represent galaxies that have their redshift in the range
$[2.7;3.4]$, and points show galaxies out of this range. The solid contour
shows the colour criteria for LBGs at $z\sim 3$
(Eqs. \ref{eq:Steidel_contour}) , and the dashed contour shows our
optimised contour for selecting galaxies at $z\sim 3$ (section
\ref{sec:new_contour}). Notice the very steep rise in $U_n-G$ around
$z\sim 3$, mainly due to the IGM attenuation. Photometric errors are
not added in this plot.}
\label{fig:ccplot}
\end{figure}

For a galaxy at this redshift, the Lyman break falls between the
$U_n$ and $G$ filters \citep{SteidelHamilton93}. It makes
the $U_n-G$ colour of galaxies very sensitive to small variations of redshift
around this value, as shown by the very steep slope in the colour-colour diagram
of Fig. \ref{fig:ccplot}. This spread of the galaxy distribution
along the $U_n-G$ axis leads to a precise selection of galaxies
in redshift space.  The colour criteria for selecting objects at
$z\sim 3$ proposed by \citet{SteidelPettiniHamilton95}, which we adopt throughout
this paper, is the following : \beq
\label{eq:Steidel_contour}
\begin{array}{lll}
R &\leq& 25.5 \\ 
G - R &\leq& 1.2 \\ 
U_n - G &\geq& G- R +1 \\ 
U_n - G &\geq& 1.6 \\ 
G - R &\geq& 0.0.
\end{array}
\eeq

In Fig. \ref{fig:ccplot}, we show a colour-colour diagram for a mock
catalogue of 360 arcmin$^2$.  Only galaxies with an apparent AB
magnitude lower than $25.5$ in the $R$ band are shown, in order to mimic 
the detection limit quoted by these authors. We represent galaxies which
have redshifts between $2.7$ and $3.4$ as diamonds. Galaxies with
other redshifts are marked as simple dots. The colour criterion
given by Eq. \ref{eq:Steidel_contour} is shown as the solid line.

Notice that the distribution of galaxies in the colour-colour plane is
smooth in  Fig. \ref{fig:ccplot}. This behaviour stems from distributing 
galaxies in a mock 
light-cone, in contrast with previous studies based on pure semi-analytical 
models (\citet{BaughEtal98}, \citet{SomervillePrimackFaber01}), 
where galaxy properties were only computed for a discrete set of time outputs. 

\subsection{Data sets} \label{sec:data_sets}
\subsubsection{Observations}
We choose to compare our modelled LBGs to the sample presented
by Steidel and co-workers, for both the variety of results they have
published so far (e.g. \citet{GiavaliscoEtal98},
\citet{SteidelEtal99}, \citet{ShapleyEtal01}), and the
large number of sources it contains. The total area surveyed 
by these authors amounts to about a thousand arcmin$^2$, and
contains $\sim$ 1250 LBG candidates (see \citet{GiavaliscoEtal98},
table 1, for a summary of observations). The density of observed LBGs
on the sky is thus about 1.2 object per square arcminute, among which
roughly 20\% are estimated to be interlopers.  Most of the
observations were made using the three filters $U_n$, $G$, and $R$,
designed to efficiently select objects at \zz{} 
\citep{SteidelHamilton93}.

A point which is still unclear in the literature so far is the definition of interlopers. In principle, these
are galaxies which, although selected by the colour criterion, are not
at the requested redshift. In practise, it seems that the definition which is used by some authors
is somewhat less strict, and objects are only called interlopers when they
are galaxies at redshifts lower than $\sim 2.3$ or stars.  We follow suit and adopt this broader definition
but remark that since there are no foreground stars in our simulation, we do not expect to find many interlopers
(see following section).

\subsubsection{Simulations}
We use the ($U_n$, $G$, $R$) filter set from \citet{SteidelHamilton93}
to compute the apparent magnitudes of our simulated galaxies. Our
standard mock catalogue size is 1 deg$^2$, and contains sources up to $z
\sim 6.5$. We also use the same apparent magnitude detection limit, that
is, $R \leq 25.5$.

As we discuss in \momaf, a unique way to build a mock
catalogue from the outputs of our simulation does not exist, because the mock
light-cone intersects only fractions of each snapshot. We choose to
introduce a random process in the generation of cones in order to be
able to make independent realisations which are all {\it a priori}
equally plausible. Because we do not include fluctuations on scales larger than the simulated volume, 
which is about 1 degree at \zz{}, the dispersion of the number of LBGs found in
different cones will yield an under-estimate of the cosmic variance
on scales of one square degree (the size of our cones). We further point out that the mean
LBG count derived from running a large number of realisation contains the full
information available in the simulation.

\begin{figure}
\centerline{\psfig{figure=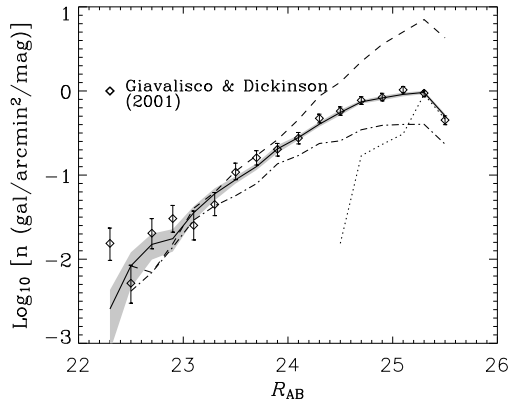,width=\hssize}}
\caption{LBG counts from \gal, compared to data from
\citet{GiavaliscoDickinson01} (diamonds with error bars). The solid
line and grey area show the mean and 1-$\sigma$ region obtained from
20 mock catalogues of 1 square degree in which we applied the maturity age selection with $t_{mat}=1.1$Gyr. The dashed line shows the LBG number counts for a
single cone, without any maturity selection, and the dot-dashed line
shows the same counts with $t_{mat}=1.3$ Gyr.
The dotted line shows counts from a higher resolution simulation 
covering a much smaller volume. Its normalisation justifies the value 
given to $t_{mat}$ (see text for discussion).}
\label{fig:counts}
\end{figure}

We ran 20 different light-cones, and found a mean number of LBGs
$<N_{LBGs}> = 4133$ per square degree, with a standard deviation of
$\sigma_{N_{LBGs}} = 183$. The modelled density of LBGs is thus $1.15$
arcmin$^{-2}$, consistent with the observed density of $1.22 \pm 0.18$
arcmin$^{-2}$ (e.g. \citet{GiavaliscoDickinson01}). In Fig.
\ref{fig:counts}, we show the mean number counts, obtained from our 20
mock catalogues (solid line), as well as the dispersion, given at
1-$\sigma$ by the grey area. Our counts compare nicely with the data
from \citet{GiavaliscoDickinson01}, over-plotted as diamonds. Note
also that, as in real observations, we find very few interlopers (about
0.1\%) with $z < 2.3$. The dashed line shows counts for our modelled
LBGs when we do not apply any maturity selection, and focus on a single example cone. 
The difference between these counts and the solid line shows
that the history resolution mentioned in
Sec. \ref{sec:hist_resolution} mainly and strongly affects the faint
end of the luminosity function (LF). Our most robust results thus
concern the bright end of the modelled LBGs, where resolution does
not affect galaxy properties. The dot-dashed line shows the counts for
the LBGs which are older than 1.3 Gyr. The dotted line shows the
counts obtained from a mock observing cone of
0.09 square degrees built with a higher-resolution $\Lambda$CDM
simulation. This higher-resolution simulation contains $320^3$ particles in
a cubic volume of side $32h^{-1}$Mpc. It has identical
cosmological parameters as given in Sec. \ref{sec:dm_sim}, and the \gal{}
post-processing was done using the same astrophysical parameter set as
in \hatton. The smaller volume and larger number of particles result in a
galaxy mass resolution $M_{res}=3.77 \times 10^8 M_\odot$, {\em i.e.} $\sim 60$ times
better resolution than in the fiducial simulation discussed here. As expected, the introduction 
of a maturity age $t_{mat}$ is irrelevant for the LBG counts of this high-resolution simulation.
 The faint-end normalisation of the counts
supports our choice of $t_{mat}=1.1$ Gyr for our maturity criterion. The
discrepancy in the bright counts are understood in terms of volume limit of the
high-resolution simulation~: just from Poisson statistics we expect about 30 times less 
objects per bin of luminosity since the volume is reduced by such a factor compared 
to the fiducial simulation. Furthermore, we have to account for cosmic variance which will affect 
more dramatically objects that are more clustered (see Sec.~\ref{sec:cosmic_variance}). 
Therefore, we cannot consider our high-resolution simulation 
as representative of the Universe as a whole, and for this reason, we prefer 
to use it
only to fix the maturity criterion in the larger, low-resolution
simulation, and we take this latter to make extensive predictions.

In the remainder of the paper, except where mentioned, we will use a single
realisation of a mock catalogue. Since we have no information on how
the data might be biased (cosmic variance), we can choose any of
our available cones. We settle for one which contains $N_{LBGs} = 4156$, in
good agreement with the data, and close to our mean value.


\subsection{Efficiency of the selection} \label{sec:efficiency}
Internal properties of galaxies are not fully eclipsed
by the IGM attenuation, so one cannot expect to extract all
the galaxies at a given redshift, and only these, from a simple region 
selection in the colour-colour plane.  It is therefore interesting to define an
efficiency of the selection process, which should account for the
fraction of lost galaxies -- that is, observable galaxies with the
right redshift which do not meet the colour criteria -- and the
fraction of interlopers restricted here to galaxies lying outside the redshift
range $[2.7;3.4]$.  We therefore define the two following quantities:
\begin{itemize} 
\item the {\it completeness} ($C$), which is the ratio of the number of selected
galaxies with redshift $\sim 3$ to the number of detectable galaxies
(i.e. galaxies with $R \leq 25.5$) at $z\sim 3$.  
\item the {\it confirmation rate} ($CR$) which is the ratio of the number of selected galaxies
with $z\sim 3$ to the total number of selected galaxies (with any redshift).
\end{itemize}
With these two complementary quantities, we can define a measurement
of the efficiency of a selection process as $E = C \times CR$. This
quantity takes values from $0$ to $1$, $0$ meaning no source was
detected at the expected redshift, and $1$ meaning that all the detectable
sources with $z \in [2.7;3.4]$ were selected, and only these.
For our sample of galaxies extracted from a 1 square degree field, we find the
following values for the colour criteria given by Steidel and co-workers (eqs
\ref{eq:Steidel_contour}) : $C = 96$\%, $CR = 88$\%, and $E \sim 85$\%.
This high efficiency shows that the LBG selection is indeed a very
good method for selecting distant galaxies, and the high completeness
shows that this selection grabs 96\% of the detectable galaxies at
\zz.

In an attempt to mimic observations, we also added photometric errors
to our magnitudes. These are simply modelled here as
uniformly-distributed random values, between $\pm \delta$ mag, in the
three bands $U, G, R$.  The selection efficiency drops to $E=70\%$ when $\delta$
reaches $0.25$ mags. For a more conservative error amplitude of $0.1$
magnitude, the efficiency settles around $80$\%.  Obviously,
photometric errors are far more complex than this simple model and
should take into account systematics. Nevertheless, their impact on LBG selection 
as assessed by our simple estimate constitutes a useful first estimate. 

\subsection{New contour} \label{sec:new_contour}
Until now, the colour selection used to select distant galaxies from
multi-band imaging observations has been defined by placing
synthetic galaxies, with a 'wide range' of ages, star formation
histories, metallicities, dust content and redshifts
(e.g. \citet{SteidelPettiniHamilton95}, \citet{MadauEtal96}), onto a
colour-colour plane, and by finding out which region encloses most distant
galaxies and fewest interlopers. This method, although spectroscopic
follow-up showed it gives good results, sweeps under the carpet the
question of the nature of the distant objects. It is expected to be
robust anyway since IGM extinction --which is external to galaxies--
plays a dominant role in the location of galaxies in the colour-colour
plane. However, one might wonder whether a tighter set of criteria could be
found if the 'wide range' of properties was replaced by more 'plausible'
properties for galaxies at a given redshift.  We present here the
first attempt to build such criteria with an {\it ab initio} model of
galaxy formation, in which distant galaxies' properties come out
naturally of the hierarchical evolution within the dark matter content of the
universe.

We found our best contour in the colour-colour plane using the {\it downhill
simplex method} \citep{NumericalRecipes} in order to maximise the efficiency
of the LBG-selection for a pentagonal contour with two sides bound to
be vertical, and one horizontal in the $(U_n-G;G-R)$ plane. For this
maximisation, we used our full fiducial catalogue of one square degree
so as to have a significant number of LBGs.  The contour we calculate is
the following :
\beq 
\nonumber
U_n-G & \geq & 1.7 ,\\ 
\nonumber G-R & \geq & -0.2 ,\\
\nonumber G-R &\leq & 1.9 ,\\
\nonumber U_n-G & \geq & 0.9(G-R)+1.1 ,\\
\nonumber U_n-G& \leq & 0.6(G-R)+5.1 .
\eeq

This contour (the dashed lines of Fig. \ref{fig:ccplot}) gives an
efficiency $E = 88$\%, instead of $E=85$\%
for the contour defined by Eq~\ref{eq:Steidel_contour}. Our efficiency is higher because we
discard a large fraction of high-$z$ interlopers by setting an upper
limit on the $U_n-G$ colour, which increases our confirmation rate to
$CR = 95$\%. However,
our contour decreases the completeness ($C=94$\%), precisely because
of this more accurate selection.  In practise, the aim of LBG surveys
is twofold : (i) select a large number of high-$z$ candidates, with
redshifts as high as possible, allowing for a broad redshift range, and
(ii) use the detected galaxies to describe the properties of 'normal' galaxies at a
given redshift.  In case (i), one wishes to favour completeness
relative to confirmation rate. In the second case, one wishes to reach
a compromise between $C$ and $CR$. Bearing this in mind, it is
worth pointing out that our selection is quite similar to that used by
Steidel and collaborators. This gives the overall impression that the
UV-dropout technique is quite robust, and independent of stellar
population modelling. As mentioned earlier, this robustness was expected since the
locus of galaxies in the colour-colour plane is mainly given by IGM
absorption, and thus by redshift.

\subsection{Cosmic variance} \label{sec:cosmic_variance}

\begin{table*}
\begin{center}
\begin{tabular}{lccccc} 
\hline\hline 
LBGs          & 4 arcmin$^2$       & 16 arcmin$^2$      & 64 arcmin$^2$      & 400 arcmin$^2$     & 1600 arcmin$^2$ \\
              &$<N>$ ($\sigma_{N}$)&$<N>$ ($\sigma_{N}$)&$<N>$ ($\sigma_{N}$)&$<N>$ ($\sigma_{N}$)&$<N>$ ($\sigma_{N}$)\\
\hline
$R \leq 25.5$ & 4.62 (2.76)        & 18.35 (7.07)       & 73.00 (18.9)       & 454.6 (66.9)       & 1817  (154)  \\
$R \leq 25. $ & 2.71 (1.94)        & 10.74 (4.80)       & 42.74 (12.2)       & 266.1 (42.1)       & 1064  (93.9) \\
$R \leq 24.5$ & 1.21 (1.20)        & 4.78  (2.74)       & 19.04 (6.69)       & 118.3 (21.4)       & 473.1 (46.6) \\
\hline
\end{tabular}
\end{center}
\caption{Mean expected numbers of LBGs brighter than $R=25.5$, $25$ and
$24.5$, in fields of 4, 16, 64, 400 and 1600 arcmin$^2$. The
standard deviations are also given between brackets. These standard
deviations are systematically higher than the square-root of the
number of sources, because of the strong effect of clustering.}
\label{tab:cosmic_variance}
\end{table*}

The degree to which observations of small areas are representative of
the whole Universe is a crucial question when one wishes to interpret
data in the cosmological context. Since our model contains spatial
information, it is possible to study the cosmic variance of LBG
densities, at least to a certain extent, because our simulation is
volume-limited and does not include long-wavelength fluctuations.

We estimated the cosmic variance as the field-to-field variation in
the number of LBGs in sub-fields, extracted from our 20 different 1
deg$^2$ fields. For each size of sub-fields, ranging from 4 arcmin$^2$
to 1600 arcmin$^2$, we used 20000 random sub-catalogues (1000 per 1
deg$^2$ mock catalogue) to estimate the mean number of LBGs and the
associated variance.

In table \ref{tab:cosmic_variance} we give the mean numbers of LBGs
and the corresponding standard deviations for fields of 4, 16, 64,400
and 1600 arcmin$^2$. Numbers are given as well for LBGs brighter than
$R=25$ and $R=24.5$.  The fact that the standard deviation is higher
by a factor $> 2$ to the poissonian deviation is a direct signature of
the high clustering properties of LBGs.

\begin{figure}
\centerline{\psfig{figure=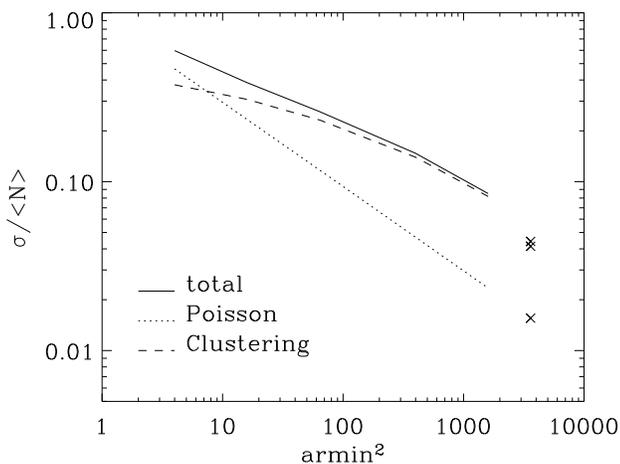,width=\hssize}}
\caption{Standard deviation in number counts of LBGs as a function of
the survey area in square arcminutes. The
solid curve is the measured standard deviation, and the dotted one is the
Poissonian prediction. The dashed curve shows the quadratic difference
between the measured deviation and the Poissonian one, namely, the
contribution of clustering to the cosmic variance.}
\label{fig:cosmic_variance2} 
\end{figure}

In Fig. \ref{fig:cosmic_variance2} we show the variation of the
relative deviation ($\sigma / \langle N \rangle$) versus the surface
of the mock survey. The continuous line shows the measured deviation
and the dotted line shows the Poissonian prediction. The dashed line
shows the quadratic difference between those : $\sigma_c / \langle N
\rangle = (\sigma^2_N - \sigma^2_{Poisson})^{1/2} / \langle N
\rangle$, where $\sigma_c$ is the contribution of clustering to the
cosmic variance. The three crosses at 3600 arcmin$^2$ show the total
dispersion and the contributions of clustering and Poisson, from top
to bottom. These were estimated only from our 20 mock catalogues and
thus rely on less statistics than the curves.  On small
scales ($<7$ arcmin$^2$), where the sources are rare enough, the
Poissonian noise dominates the variance. On larger scales, the
variance is dominated by the clustering of LBGs, as the Poissonian
variance falls off more rapidly.  As the Universe is homogeneous on
large scales, we expect the clustering contribution to dwindle
progressively as the size of the survey increases. This regime is however
out of reach with our current simulation because the simulated volume
has an angular size of $\sim1$ square degree at \zz{}. Above this
scale we are bound to under-estimate the dispersion. Because of this,
we cannot tell at the moment whether the points at 3600 arcmin$^2$ in
Fig. \ref{fig:cosmic_variance2} are low due to finite volume effect or
to the fact that we reached the point where clustering strength starts
to vanish.

\section{Clustering}
The clustering properties of LBGs give important constraints on the
dynamical processes that drive galaxy formation. Moreover, they can be
observed directly and no assumptions are necessary to compute the
angular correlation function. In this section, we show different
aspects of the clustering properties of our modelled LBGs, namely, 
halo masses, the location of LBGs in the cosmological simulation, the
angular correlation function, the spatial correlation function, and
the bias of LBGs relative to dark matter.

\subsection{Halo masses}
\begin{figure}
\centerline{\psfig{figure=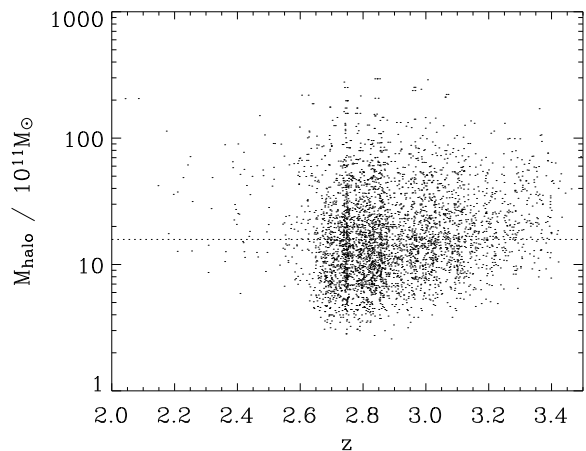,width=\hssize}}
\caption{Masses of LBG host haloes as a function of redshift. The
horizontal dotted line shows the median mass value of $\sim 16$
$10^{11}M_{\odot}$.}
\label{fig:halo_mass}
\end{figure}

A first impression of the clustering properties of LBGs can be given
by the mass distribution of the dark matter haloes which harbour
them. In Fig. \ref{fig:halo_mass}, we show the masses of these haloes
versus the redshifts at which the LBGs are detected in our
simulation. The median halo mass for our LBGs is $15.8\times
10^{11}M_{\odot}$, as indicated by the horizontal dotted line. This
mass is about ten times the halo mass resolution of
the simulation. The median values for populations of modelled
LBGs with various luminosities are given in table \ref{tab:halo_props}. The relatively large
mass of the haloes harbouring our modelled LBGs at $z\sim 3$ indicates
that these galaxies are highly biased with respect to the dark matter
distribution. This matches qualitatively what is observed in different
surveys (\citet{GiavaliscoEtal98}, \citet{ArnoutsEtal02}).

\begin{table} 
\begin{center}
\begin{tabular}{lc} 
\hline\hline
LBGs        &    $<M_{halo}>$      \\
            &($10^{11} M_{\odot}$) \\
\hline
$R\leq25.5$ & 15.8      \\
$R\leq25$   & 15.4      \\
$R\leq24.5$ & 14.0      \\
\hline
\end{tabular}
\end{center}
\caption{Median masses of DM haloes harbouring LBGs.}
\label{tab:halo_props}
\end{table}

\subsection{Location of LBGs and their descendants}
\begin{figure*}
\begin{center}
\begin{tabular}{cc}
\psfig{figure=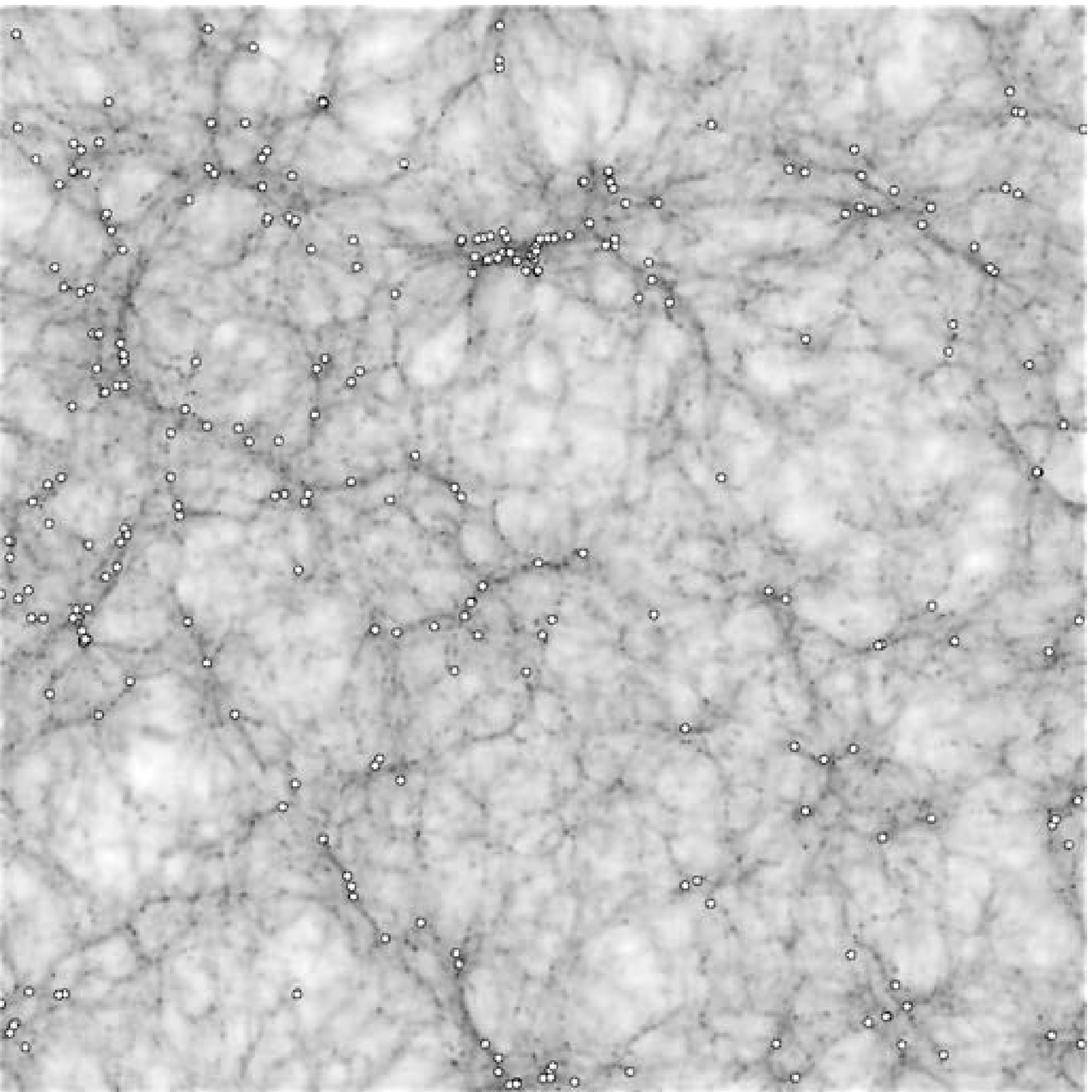,width=\hssize} & 
\psfig{figure=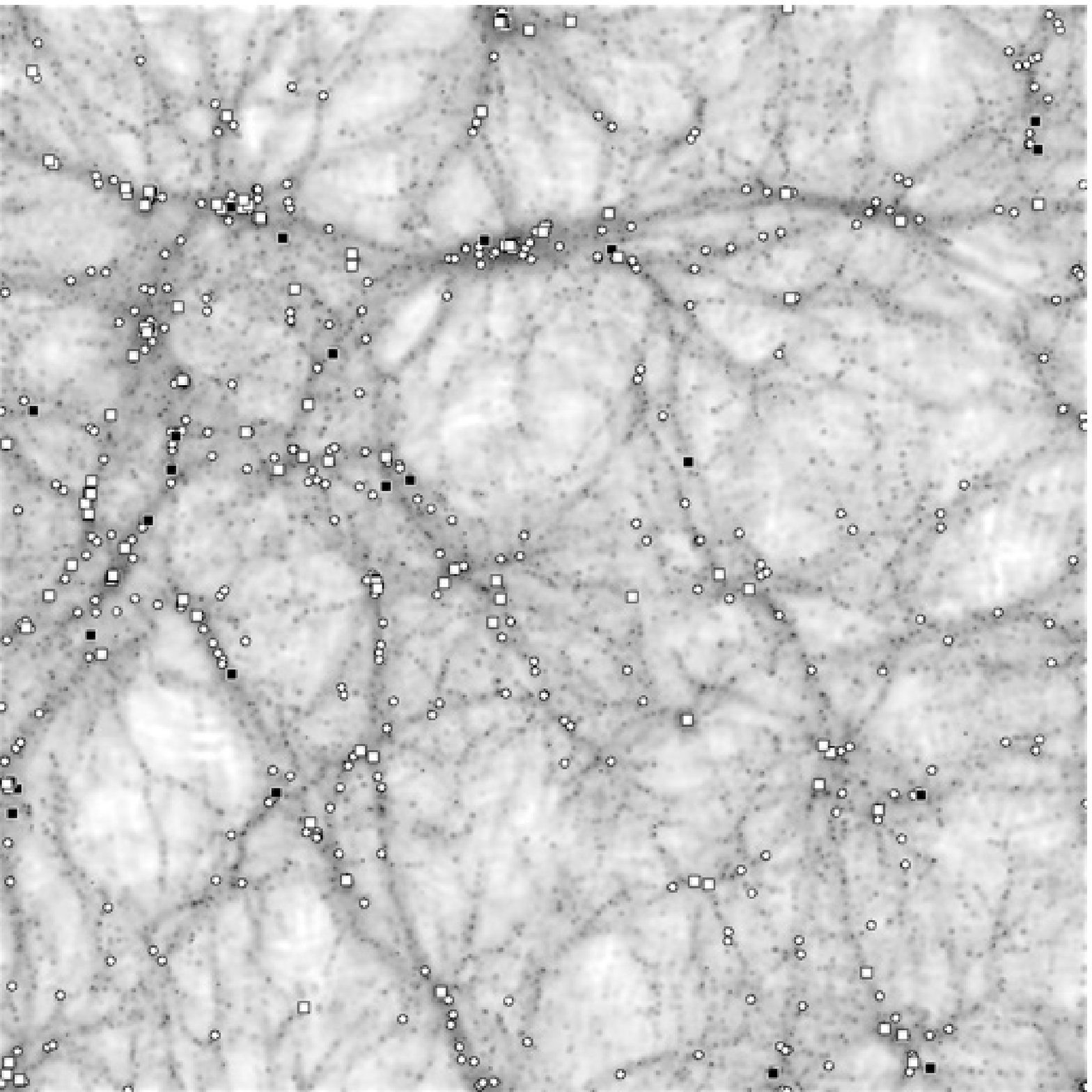,width=\hssize}
\end{tabular}
\end{center}
\caption{{\it Left hand panel} : location of LBGs at z=3. White discs show
LBGs, and the DM density (in log) is shown in grey scale, the darker the
denser.  {\it Right hand panel} : location of local ellipticals and the
descendants of LBGs in the same comoving volume, at redshift 0. The
white symbols show elliptical or lenticular galaxies, the squares
indicate descendants of LBGs that are present in the volume of the left hand
side panel. Black squares show spiral descendants of LBGs.  The LBGs
and their descendants are highly clustered, located in dense DM
regions. The projected volume shown is 150 Mpc on a side, and has a
depth of 15 Mpc.}
\label{fig:LBGSlice}	
\end{figure*}

On the left-hand side panel of Fig. \ref{fig:LBGSlice}, we show the
location of LBGs at $z\sim 3$ in a slice of our simulated cosmological
volume. The white discs show LBGs, and the underlying grey background
shows the dark matter density (the darker the denser), in
logarithmic scale. The LBGs are clearly highly clustered in this
picture. The right-hand panel shows the location of the descendants of
LBGs at $z=0$ as squares. Black squares indicating spiral galaxies, and
white squares elliptical or lenticular galaxies. The white discs are
elliptical/lenticular galaxies which do not have a LBG progenitor at
\zz. The descendants of LBGs are even more highly clustered than LBGs
themselves, as they fell into DM structures from $z=3$ to $z=0$. We will come
back to the properties of the descendants of LBGs in section
\ref{sec:descendants}.

\subsection{Angular correlation function}\label{sec:ang_cor}

For the computation of the angular correlation function (ACF), we
used the estimator proposed by \citet{LandySzalay93}
(hereafter LS93) : 
\beq \label{LS1993} 
w(\theta) = \frac{DD(\theta) - 2DR(\theta) + RR(\theta)}{RR(\theta)}, 
\eeq 
where $DD(\theta)$ is the number of pairs of LBGs with
angular separation between $\theta$ and $\theta + \delta \theta$,
$RR(\theta)$ is the analogous quantity for a random catalogue, and
$DR(\theta)$ the number of observed-random pairs.

The estimate of the errors in the general case is a difficult
task. Here, we consider Poissonian noise (in the number of pairs)
only. We expect this to be a good approximation because our survey is
large enough (compared to the scales we probe), and we deal with a rather diluted
distribution of galaxies so that finite-volume errors will
be negligible compared to discreteness errors. Moreover,
considering only Poissonian error-bars implies that we underestimate
the real uncertainties. It is not a concern here as our
goal is to show that our spatial distribution of LBGs is
consistent with the observed one -- and this is the case within
Poissonian uncertainties (see next section).  Following LS93, we
write Poissonian errors as : 
\beq \label{eq:sig_ls}
\sigma_w (\theta) =
\frac{1+w(\theta)}{\sqrt{N_g(N_g-1)}} \left[ \frac{N_r(N_r-1)}{RR} \right]^{1/2}, 
\eeq 
where $N_g$ is the number of galaxies in our mock
survey, and $N_r$ the number of objects in our random catalogue
(here, $N_r \sim 7\times 10^4$). The first factor in the above equation
shows the Poissonian contribution, as it is inversely proportional to the
square root of the number of LBG pairs. The second term is a
geometrical contribution, and depends only on the shape of the
survey. In this expression of the standard deviation, we have assumed that
the integral constraint was negligible. Of course, the standard deviation
of Eq. \ref{eq:sig_ls} strongly depends on the angular bin-size we use to evaluate the
ACF. This appears only implicitly~: if the bin size doubles, RR will
also roughly double, and so $\sigma_w$ will be divided by a factor
$\sqrt{2}$. In order to compare our results to those of
\citet{GiavaliscoEtal98}, we adopt the same binning as these
authors.

Finally, note that the random transverse shifting of boxes every 150
comoving Mpc along the line of sight that we use to suppress
replication effects implies an underestimate of the ACF of about 10\%
at most, as detailed in \momaf{}.

\begin{figure}
\centerline{\psfig{figure=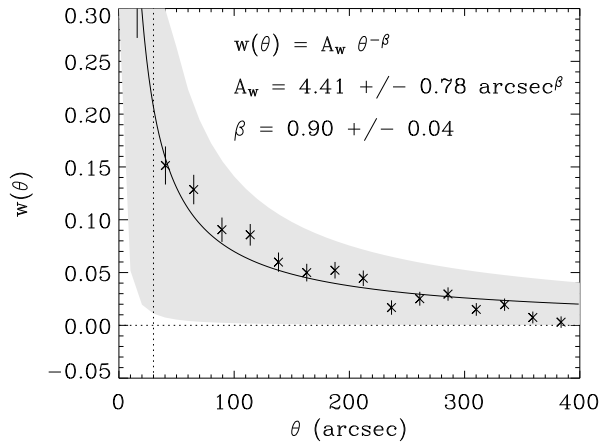,width=\hssize}}
\caption{Angular correlation function for our sample of LBGs, extracted from a 1
square degree mock catalogue.  The crosses show our estimate of
$w(\theta)$ with the LS93 estimator, and the error bars come from a
Poissonian estimate. The grey area shows data from
\citet{GiavaliscoEtal98}. This area corresponds to their ``LS-random''
estimate, which is similar to ours. The solid line
shows a maximum-likelihood fit to our data by a function $w(\theta) =
A_w \theta^{-\beta}$, with best fit values indicated in the panel.}
\label{fig:wtheta}
\end{figure}

Our measured ACF is represented in Fig. \ref{fig:wtheta}, by crosses
and their error-bars. The solid line shows a maximum likely-hood fit of
a power-law $w(\theta)=A_w\theta^{-\beta}$ to our data. The best fit
parameters are : $A_w = 4.41 \pm 0.78$ and $\beta = 0.9 \pm
0.04$. This is consistent with the results from
\citet{GiavaliscoEtal98}. Their fit to observed data is shown in
Fig. \ref{fig:wtheta} by the shaded region\footnote{The values from
these authors that we consider are those given by their so-called
``LS-random'' estimate, which corresponds to our estimate.}. The
vertical dotted line roughly shows the angular size of a cluster of
galaxies at $z\sim 3$, below which the contribution to clustering mainly comes
from pairs of galaxies within the same halos. In \gal{}, the positions of galaxies
inside halos are given according to some prescription and not directly measured from
the N-body simulation. Hence our model only makes reliable predictions for the two-halos contribution 
to $w(\theta)$.

\subsection{Spatial correlation function and linear bias}
\begin{figure}
\centerline{\psfig{figure=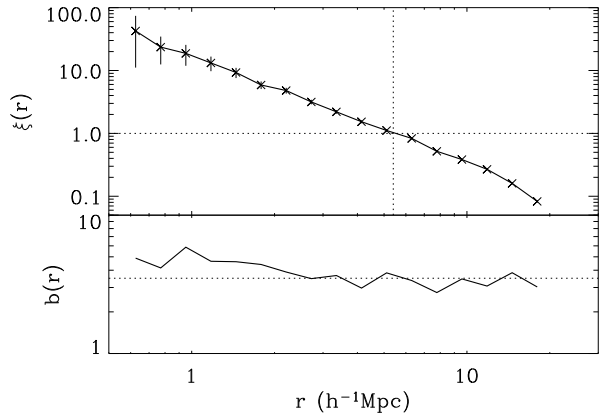,width=\hssize}}
\caption{{\it Top panel} : spatial correlation function of our modelled
LBGs in the snapshot corresponding to z=3. {\it Bottom panel} : linear
bias of our modelled LBGs with respect to the dark matter.}
\label{fig:SCF}
\end{figure}

We computed the spatial correlation function (SCF) using the same
estimator (LS93) as for the ACF. This time, however, we computed the
SCF directly from the snapshots of \gal, bypassing the cone of sight
generation. We show the result in Fig. \ref{fig:SCF}. The top panel
is the SCF of LBGs identified in the snapshot at $z=3$. The error bars
only show the poissonian deviations. The bottom panel shows the
linear bias of LBGs relative to the dark matter particles of the
cosmological N-body simulation ($b^2(r) = \xi_{LBG}(r) /
\xi_{DM}(r)$). We limit our plots to the approximate range 1-10
$h^{-1}$Mpc for two reasons : (i) we distribute galaxies inside
clusters according to a semi-analytic spherically symmetric prescription, 
so we loose angular information for some galaxies below 1
$h^{-1}$Mpc, and (ii) to avoid edge effects, we limit the study to
scales below one tenth of the simulated volume size (100 $h^{-1}$Mpc).

We find that the linear bias slowly decreases from $\sim 5$ to $\sim
3$ between 0.5 and 20 comoving $h^{-1}$Mpc.  This is compatible with
the bias estimated by \citet{GiavaliscoEtal98},
\citet{AdelbergerEtal98} and \citet{PorcianiGiavalisco02}.


\section{UV/IR luminosity budget}
One of the most important challenges toward understanding  galaxy
formation is to determine at which epoch stars formed, or, in other
words, how and when did the mass of stars we see in local Universe galaxies
assemble.  The observation of distant galaxies in the
optical window is now known to be insufficient to retrieve star
formation rates (e.g. \citet{DevriendtGuiderdoniSadat99}). The
discovery of the cosmic far infrared background, and the numerous
IR/sub-mm surveys, have made it clear that dust extinction within
galaxies plays a fundamental role in determining the UV emission of
distant galaxies, and therefore in extracting information on the
cosmic star formation rate.  The reason is that the UV luminosity of a
galaxy is governed by the amount of recent star formation --which is responsible
for young massive stars dominating the energetic part of the SED-- and
dust absorption --which is most efficient at UV wavelengths, and
re-emits light in the IR/sub-mm domain. Unfortunately, far
infrared or sub-mm instruments do not yet have sensitivity or
resolution which would make a thorough follow-up of LBGs at large wavelengths possible. 

Our model includes a treatment for dust extinction and emission in the
IR/sub-mm domains, which allows us to predict extinction
properties of LBGs, and make the connection between LBGs, seen in the
optical, and sources detected at IR/sub-mm wavelengths.  In this section,
we first discuss the UV luminosity function and selection
effects. Second, we consider the extinction properties of these
objects as predicted by \gal. Third, we show how both effects
alter the so--called Madau diagram \citep{MadauEtal96}. We eventually make
predictions concerning the detectability of the sub-mm counterparts of LBGs.

\subsection{Luminosity functions at $z \sim 3$} \label{sec:luminosity_functions}

\begin{figure}
\centerline{\psfig{figure=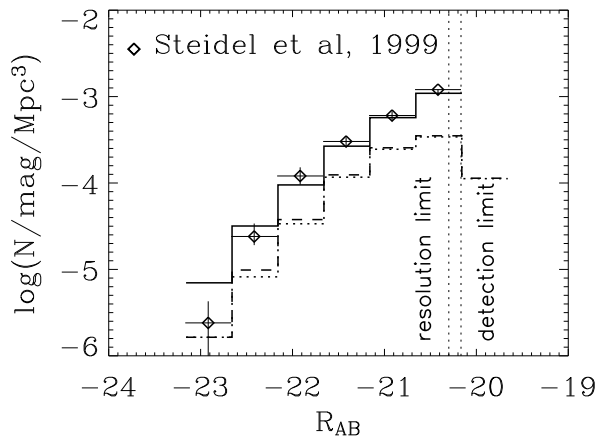,width=\hssize}}
\caption{Luminosity functions at $z\sim3$, for LBGs. The x-axis shows absolute $R$ magnitudes in the observer
frame.  The black diamonds show the data from \citet{SteidelEtal99},
converted to the current cosmology.  The solid histogram shows the LF we
compute for our sample of LBGs, using the effective volumes given in
\citet{SteidelEtal99}. We assume a single luminosity distance for
the whole sample (also given by \citet{SteidelEtal99}).  This histogram is in
fair agreement with the data. The dashed histogram shows the LF computed
for all the detectable galaxies (i.e. galaxies with $R\leq 25.5$) in
our sample that have a redshift between $2.7$ and $3.4$.  Here, we
normalised the LF to the total volume of the cone of sight between
redshifts 2.7 and 3.4. We also used the ``true'' luminosity distance
of each galaxy. This LF is lower than the previous one because of
volume corrections. Finally the dotted histogram shows the true luminosity 
function of LBGs only, {\em i.e.} without volume corrections. The two vertical dotted lines show the
resolution and detection limits.}
\label{fig:funclum1}
\end{figure}

In Fig. \ref{fig:funclum1}, we compare different estimates of our
measured luminosity function (LF) at \zz\ with the LF given by \citet{SteidelEtal99}.
\begin{itemize}
\item The {\it solid histogram} shows the LF for all our modelled
LBGs, computed in the same way \citet{SteidelEtal99} computed their observed LF. First we
divide the number of LBGs in each apparent-magnitude bin by the
effective volumes quoted by these authors for our cosmological parameters.  Then, we
convert our apparent magnitudes to absolute magnitudes using the
relation $L_{\nu}=10^{-0.4(48.6+m_{AB})} \times 4\pi d_L^2/(1+z)$
where $d_L = 5.6\ 10^{28}\ h^{-1}$cm, is the single value of the 
luminosity distance given by \citet{SteidelEtal99} for $z=3.04$ in a
$\Lambda$CDM universe ($\Omega_{\Lambda} = 0.7$, and $\Omega_{m} =
0.3$). Note that the difference in luminosities induced by using a single redshift and
luminosity distance are negligible when compared to the uncertainties in the volume corrections.
\item The {\it dashed histogram} shows the luminosity function of all 
detectable galaxies ($R\leq 25.5$) with redshifts between 2.7 and
3.4. This time, we first convert apparent magnitudes to absolute
magnitudes using each galaxy's luminosity distance and redshift. 
We then divide the number of objects in each magnitude bin by the total volume of the
cone of sight between $z=2.7$ and $z=3.4$, ignoring any volume effect
due to the selection function.
\item The {\it dotted histogram} is the same as the previous one, but only
includes LBGs. It lies just below the dashed histogram in every
bin, showing that LBG selection is more or less equally efficient at all
magnitudes.
\end{itemize}
The {\it resolution} and {\it detection} limits are shown
as dotted vertical lines. The former limit was defined in section
\ref{sec:res_effect}. The latter is given by \citet{SteidelEtal99} 
and corresponds to the apparent magnitude limit $R \leq 25.5$.  Our
resolution limit is of the same order.

The solid curve of Fig. \ref{fig:funclum1}, which reproduces the
analysis of \citet{SteidelEtal99}, yields a very encouraging match to the
data. At this stage, we remind the reader that in this paper we use the
same \gal~ model that was used in \hatton~ to reproduce local
galaxy properties. It is therefore an impressive result that we manage to match
the observed luminosity function (or the number counts shown in Fig.
\ref{fig:counts}) of galaxies at high redshift as well.

\begin{figure}
\centerline{\psfig{figure=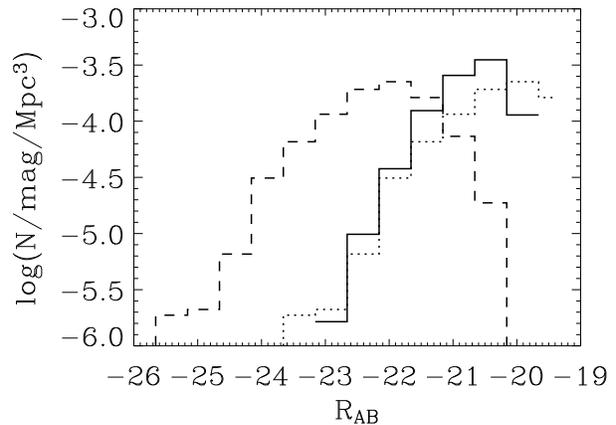,width=\hssize}}
\caption{Effect of dust extinction on the luminosity function of
galaxies at \zz.  The solid histogram shows the LF of all our modelled LBGs
at \zz, as a function of their apparent $R$ magnitudes.  The dashed
histogram is the LF these galaxies would have if they contained no dust. It
was computed directly from the rest-frame emission of galaxies at
1600\AA, assuming no extinction. Finally the dotted histogram presents the same results
as the dashed one but shifted by 2 magnitudes to facilitate comparison with 
the solid histogram.}
\label{fig:funclum3}
\end{figure}

In Fig. \ref{fig:funclum3}, we show the effect of dust extinction on
luminosity functions. The solid histogram on this figure represents the absolute $R$ magnitude (in the
observer frame) luminosity function, and includes all our modelled
galaxies at \zz\ (with the usual $R\leq 25.5$ criterion on apparent
magnitude). The dashed histogram corresponds to the luminosity function of all these
galaxies computed directly from their luminosities at 1600\AA\ in the
rest-frame, {\it assuming no extinction}. The dotted histogram is the
dashed one shifted two magnitudes fainter to facilitate
the comparison with the extinguished LF (solid histogram).  
Note that the filters differ for the two
LFs. Indeed, the $R$ filter blue-shifted to \zz\ has a width $\Delta \lambda
\sim 375$\AA\ and is centred at $\lambda_R \sim 1700$\AA, whereas the filter
we used to compute the 1600\AA\ emission is centred at this wavelength
(in the galaxy rest-frame), and is a top-hat function of width 20\AA{}.
Nevertheless the huge difference between the two histograms demonstrates how
delicate it is to extract the UV flux really emitted from observed
broad band fluxes. We emphasise that it is this very same UV flux which is mainly used 
to compute the cosmic star formation rate at intermediate and high redshift. 
Moreover, as a final note of caution, we stress that even the slopes of the LFs differ, which means that a
unique conversion factor in not sufficient to accurately correct data for extinction.

\subsection{Extinction of LBGs}\label{sec:extinction}

Obscuration by dust plays a key role for the 
extraction of  star formation rates from UV observations. Here,
we show how LBGs and other detectable galaxies at \zz\ are affected by
this process, and we give the mean extinction factors predicted by \gal\
for these galaxies.

\subsubsection{Qualitative view}
\begin{figure}
\centerline{\psfig{figure=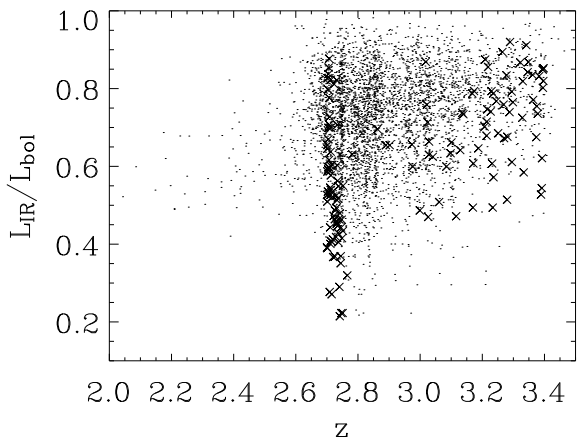,width=\hssize}}
\caption{ Ratio of IR luminosity (8$\mu$m
$< \lambda <$ 1000 $\mu$m) to bolometric luminosity for LBGs (dots)
and detectable galaxies with $z\in [2.7;3.4]$ (crosses) that do not meet the
LBG selection criteria. Most of these galaxies emit more than half their light in
the IR/sub-mm.}
\label{fig:bols_vs_z}
\end{figure}
In Fig. \ref{fig:bols_vs_z} we show an estimate of the internal
extinction versus redshift for our sample of LBGs and other galaxies
at \zz{} which are brighter than $R=25.5$ (the so-called `detectable' galaxies). Internal
extinction is estimated here as the ratio of infrared bolometric
luminosity (integrated from 8$\mu$m to 1000$\mu$m at rest) to
bolometric luminosity. A value of 0.5 of this ratio means that half 
the light emitted by stars is absorbed and re-emitted by dust. We find
that $\sim$95\% of our LBGs emit more than half their light in the
IR/sub-mm. This fraction of heavily extinguished objects holds for all
detectable galaxies at \zz. This suggests that an important
contribution of the IR/submm background is due to \zz\ galaxies.

The fact that missed galaxies (crosses in Fig. \ref{fig:bols_vs_z})
lie on rather narrow redshift bands on either side of $z\sim 3$  
is another manifestation of the LBG selection efficiency: foreground galaxies are
only missed because of border effects in the colour-colour plane, whereas background 
galaxies are more extinguished on average.

\subsubsection{Extinction factor}
The usual way to quantify extinction is through the absorption
coefficient $A(\lambda) = k(\lambda) E(B-V)$, where $k(\lambda)$ is
the adopted attenuation law. In terms of luminosities, this
coefficient can be written as \beq A(\lambda) = 2.5 \log \frac{L_{no\
ext}(\lambda)}{L_{ext}(\lambda)}, \eeq where $L_{no\ ext}(\lambda)$ is
the non-extinguished luminosity at wavelength $\lambda$, and
$L_{ext}(\lambda)$ the extinguished one, computed for a given
inclination of the galaxy. The {\it extinction factor} is then defined
by $F(\lambda) = 10^{0.4 A(\lambda)}$, that is, in terms of
luminosities, $F(\lambda) = L_{no\ ext}(\lambda) /
L_{ext}(\lambda)$. This quantity is of great importance because it is
necessary for the extraction of information about the star formation
rates of distant galaxies. However, its value is still a matter of
debate. This is partly due to averaging effects, as emphasised by
\citet{MassarottiIovinoBuzzoni01}. Indeed these authors stress that
computing an {\it average} extinction factor from an average
absorption coefficient -- namely \beq F_{ave}(\lambda) = 10^{0.4
\sum_i A_i(\lambda)/N} = 10^{0.4 <A(\lambda)>}, \eeq where $N$ is the
number of galaxies in the sample -- leads to an underestimate of the
effect of dust. They suggest that one should rather use an {\it
effective} extinction factor defined by \beq F_{eff}(\lambda) =
\frac{1}{N} \sum_i 10^{0.4 A_i(\lambda)} = 10^{0.4 A_{eff}(\lambda)}.
\eeq In order to understand the difference between both estimates,
it is easier to re-write them in terms of luminosities. The
{\it average} extinction factor becomes the geometric mean \beq \label{Fave}
F_{ave}(\lambda) = \left[ \prod_i \frac{L_{no\
ext}^i(\lambda)}{L_{ext}^i(\lambda)} \right]^{1/N}, \eeq whereas the
{\it effective} extinction factor is simply the arithmetic mean \beq \label{Feff}
F_{eff}(\lambda) = \frac{1}{N} \sum_i \frac{L_{no\
ext}^i(\lambda)}{L_{ext}^i(\lambda)}.  \eeq We have computed both factors for our modelled
galaxies, in order to facilitate the comparison with \citet{SteidelEtal99}
who use $F_{ave}(1600) = 4.7$, and also to see whether they give us results
which differ by a factor $\sim$ 3, as claimed by
\citet{MassarottiIovinoBuzzoni01}. We summarise our results in table
\ref{tab:ext_factors}. We only find a discrepancy of about a factor
1.3 between the two estimators, with extinction factors
$F_{eff} \sim 6.2$ and $F_{ave} \sim
4.75$. These values are much lower than the value of 12$\pm$2 obtained by
Massarroti and co-workers, but their analysis is based on the Hubble
Deep Field only, and includes galaxies spanning a much wider range of
redshifts than those we consider here.  Our {\it average} extinction factor has a
value close to that used by \citet{SteidelEtal99}, and our {\it effective}
extinction factor is consistent with the detailed analysis of
\citet{AdelbergerSteidel00} who conclude the effective extinction should be around a factor $\sim 8$.

\begin{table} 
\begin{center}
\begin{tabular}{l|cc}
\hline 
\hline
LBGs            &  $F_{eff}(1600)$  &  $F_{ave}(1600)$  \\
\hline 
$R\leq 25.5$    &      6.19         &      4.75         \\
$R\leq 25$      &      6.13         &      4.67         \\
$R\leq 24.5$    &      6.21         &      4.83         \\
\hline
\end{tabular}
\end{center}
\caption{Extinction factors at 1600 \AA{} for different sub-populations of modelled LBGs.
The two estimators described by Massarotti et al (2001) are
given for each sub-sample.}
\label{tab:ext_factors}
\end{table}

\subsection{Cosmic star formation rate}

Another uncertain step in retrieving star formation rates from UV
fluxes is the UV-SFR connection, considering no absorption (or no
dust). This relation is quite sensitive to the initial mass function
(IMF) and we remind the reader that the IMF in this analysis is the same we used in our fiducial 
model described in \hatton, namely a \citet{Kennicutt83} IMF from 0.1 M$_{\odot}$ to 120 M$_{\odot}$.

\subsubsection{UV/SFR conversion}
The first step is to obtain extinction factors, in order to relate
the observed magnitudes to the UV flux that galaxies emit prior to
extinction. The second step is to make the link between these
extinction-corrected UV luminosities and star formation rates. This effectively
is feasible because massive stars are the main contributors to the galaxy
UV flux and are sufficiently short-lived. However, star formation rates have
to be averaged over a period of about 100 Myr so that a robust enough
relation may be establish (e.g. \citet{Kennicutt98a}). In table
\ref{tab:UV_sfr_coversion} we give SFR-to-UV luminosity ratios for different
sub-populations of LBGs, whose SFRs are evaluated as the mass of
stars younger than 100 Myr, divided by that period of time. Because of the rapid timescale 
of stellar evolution, this is an
under-estimate of the real SFR, by about 20\% for Kennicutt IMF. We also
give ratios computed using the instantaneous SFR for
comparison. Both these estimates yield reasonable values because the evolution of the
SFR is generally small over a 100 Myr period and therefore 
are consistent with \citet{MadauEtal96} -- who use a conversion factor
of $9.54$ M$_{\odot}$ yr$^{-1}$ per
$10^{29}$erg s$^{-1}$ Hz$^{-1}$-- and with Kennicutt (1998), who
gives a value of $14$ M$_{\odot}$ yr$^{-1}$ per
$10^{29}$erg s$^{-1}$ Hz$^{-1}$. The ratios we compute, however, are the result of 
particular star formation histories: those of our LBGs. Looking at table~\ref{tab:UV_sfr_coversion}, 
one remarks that averaged SFRs are systematically larger than their instantaneous counterparts by about
a factor two. This indicates that modelled LBGs tend to be identified as such when they enter a ``post-starburst'' phase.
\begin{table} 
\begin{center}
\begin{tabular}{l|c c}
\hline
\hline
LBGs            &  SFR$_{100}/$ L(1600) & SFR$_{inst}/$ L(1600)  \\
\hline
$R\leq 25.5$    &   16.7    (3.3)       & 8.94  (2.2)\\
$R\leq 25$      &   16.8    (3.6)       & 8.88  (1.1)\\
$R\leq 24.5$    &   16.8    (4.5)       & 8.77  (0.9)\\
\hline
\end{tabular}
\end{center}
\caption{SFR to UV luminosity ratios (in M$_{\odot}$ yr$^{-1}$ per
$10^{29}$erg s$^{-1}$ Hz$^{-1}$) computed for different subsamples of
our modelled LBGs. The UV luminosities are computed as {\it if the
galaxies contained no dust}, that is, with no extinction. In the first
column, the SFR is computed as the mass of stars younger than 100 Myr
divided by 100 Myr. In the second column, the SFR is the instantaneous
star formation rate, computed over a Myr only. For sake of completeness we
also give standard deviations between brackets for each sub-class of LBGs.}
\label{tab:UV_sfr_coversion}
\end{table}

\subsubsection{The cosmic star formation rate history}
Following \citet{SteidelEtal99}, we compute the cosmic SFR at \zz{} from our UV luminosity function.
Integrating the modelled UV LF of LBGs and converting
it into a star formation rate density using Table \ref{tab:UV_sfr_coversion}, we find 
$\rho_{SFR} = 8.9 \times 10^{-3}$ M$_\odot$yr$^{-1}$Mpc$^{-3}$. Correcting this value for extinction
with the effective factor of Table \ref{tab:ext_factors} yields a density $\rho_{SFR} = 5.5 \times 10^{-2}$ 
M$_\odot$yr$^{-1}$Mpc$^{-3}$. The last step is to correct for the missing faint end of the luminosity
function (unresolved galaxies). Following \citet{SteidelEtal99}, we apply a correction of 0.5 dex and find 
$\rho_{SFR} = 0.17$ M$_\odot$yr$^{-1}$Mpc$^{-3}$, which is consistent with \citet{SteidelEtal99} and 
\citet{AdelbergerSteidel00}.

\subsection{Prediction of sub-mm flux}

One of the main ambitions and major difficulties of hierarchical
models of galaxy formation is to reproduce the sub-mm number counts.
Several attempts have been made in semi-analytic
models to fit these observations, but none of the recipes has yet
emerged as a well accepted solution. \citet{GuiderdoniEtal98}
implemented a massive IMF to obtain SCUBA objects, but there is neither a
real physical motivation for this solution, nor strong
observational hints. \citet{DevriendtGuiderdoni00} introduced an ad-hoc
redshift evolution which did not possess firmer 
foundations. Finally, \citet{BallandDevriendtSilk03} used a simple yet
physically motivated model to estimate energy exchanges during galaxy
interactions which they propose as the principal mechanism for triggering 
starbursts. They show that this solution yields a good match to mid-IR
as well as SCUBA counts. A thorough discussion on this topic will be given
in paper 4, but we forewarn the reader that the fiducial \gal{} model presented in
\hatton{} that we are using for the present study probably underestimates 
the cosmic amount of 850 $\mu$m light by a factor $\sim$2.

\begin{figure}
\centerline{\psfig{figure=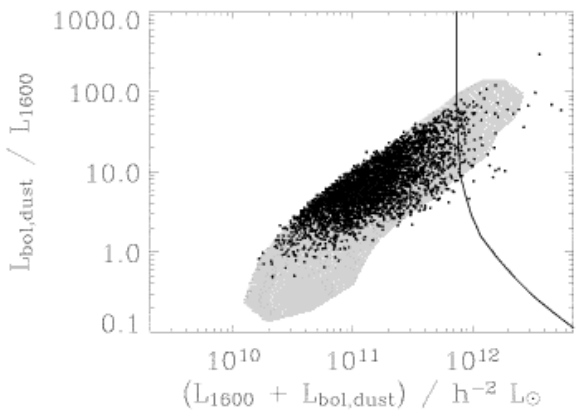,width=\hssize}}
\caption{Ratio of IR to 1600\AA\ luminosity versus
the sum of these luminosities. In a first approximation, this shows
absorption by dust versus star formation rate. The grey region
roughly indicates the location of observed LBGs
\citep{AdelbergerSteidel00}, whereas the dots represent our modelled LBGs.
The solid line marks the SCUBA detection limit at 850 $\mu$m.}
\label{fig:adelberger_special}
\end{figure}

In Fig. \ref{fig:adelberger_special} we show ratios of IR 
to UV (1600\AA) luminosities versus the sum of these luminosities
for our modelled LBGs. As a crude approximation, this is equivalent to
``dust absorption'' versus ``star formation rate''.  The grey regions broadly
indicate the contour of the data analysed by
\citet{AdelbergerSteidel00}, and the solid line shows the SCUBA
detection limit.  The good match between observations and our modelled
LBGs suggests that we manage to compute dust absorption and star
formation rates in a consistent way. We also recover the observed correlation
between star formation and dust content of the galaxies : the more
obscured, the more star-forming.  However, we predict that only about 1 \% 
of our modelled LBGs should be detectable with SCUBA at 850 $\mu$m 
at the 2 mJy level, in agreement with the estimates of \citet{ChapmanEtal00}.

The model can be used to make predictions for other IR/submm surveys. For 
instance, we find 
that the SWIRE\footnote{{\tt http://www.ipac.caltech.edu/SWIRE/}}
survey with SIRTF will be able to detect only 0.04\% (resp. 0\%, 0\%,
0\%, 0.02\%, 1.4\%, 2.3\%) of them at 170 $\mu$m (resp. 70$\mu$m, 24$\mu$m,
8$\mu$m, 5.8$\mu$m,4.5$\mu$m, 3.6$\mu$m) at the 17.5 mJy level
(resp. 2.75mJy, 0.45mJy, 32.5$\mu$Jy, 27.5$\mu$Jy, 9.7,$\mu$Jy
7.3$\mu$Jy). These objects are clearly a key target for ALMA which
should be able to detect all of them at 850 $\mu$m or $1.3$ mm, at the
0.1 mJy level. 

Predictions at other optical and IR/submm wavelengths can be easily retrieved 
from queries to our web interfaced relational database located at 
{\tt http://galics.iap.fr}.


\section{Nature of LBGs} \label{sec:nature_of_lbgs}
As emphasised by \citet{ShapleyEtal01}, the question of the nature of
LBGs remains quite open.  Different models suggest that LBGs are seen
because they are undergoing strong starbursts due to major mergers,
whereas others claim that LBGs are central galaxies of massive haloes
that form stars steadily so as to reach a consequent size by redshift
3. The increasing number of observations seems to indicate that LBGs
have many facets. Each of the two above-mentioned scenarios can
explain some of their properties. We describe in this section the
general properties of our modelled LBGs and investigate their
star formation histories.

\subsection{Physical properties}

\begin{figure*}
\begin{center}
\begin{tabular}{cc}
\psfig{figure=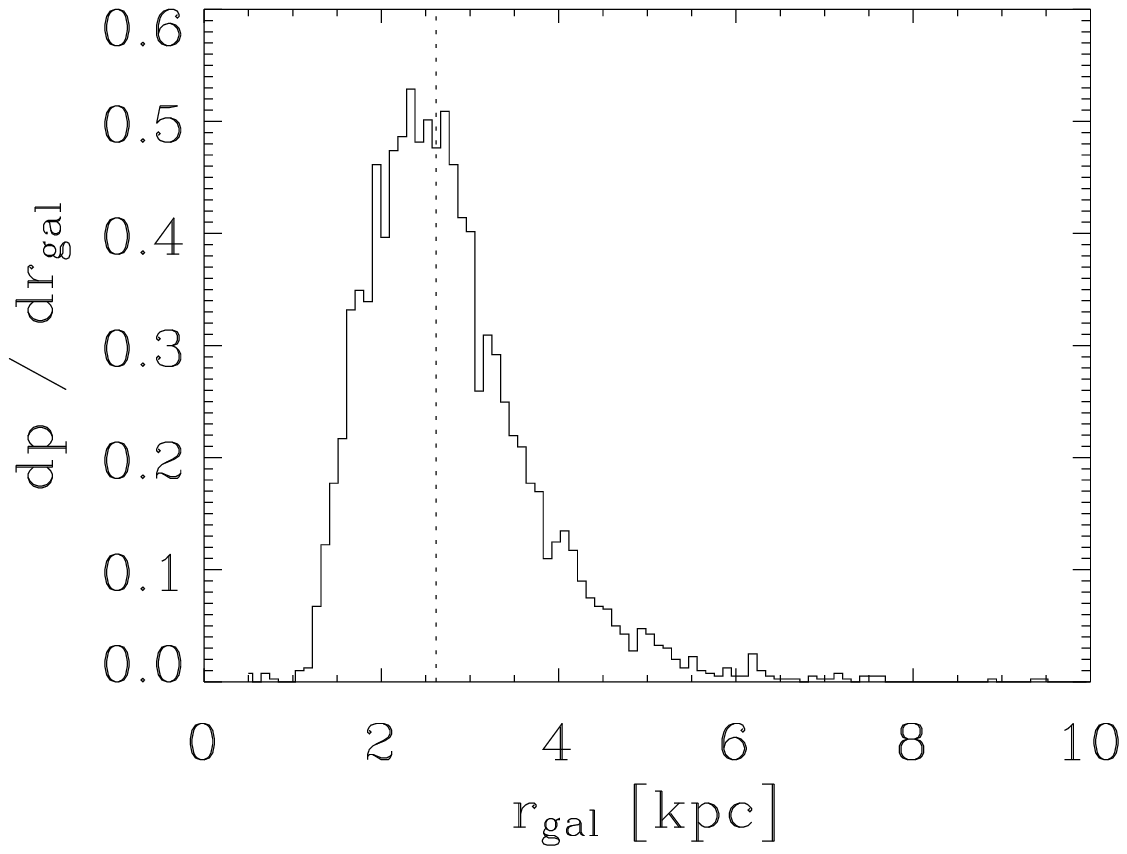,width=\hssize} & 
\psfig{figure=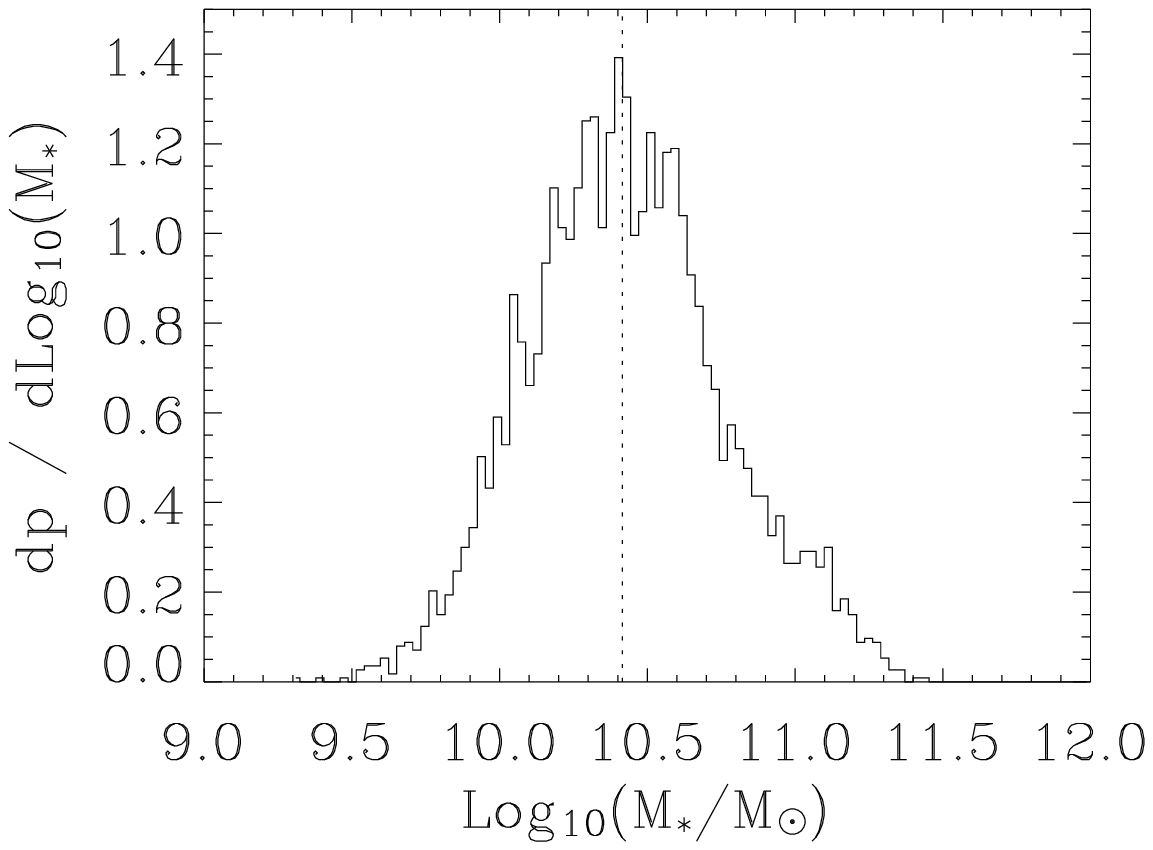,width=\hssize} \\
\psfig{figure=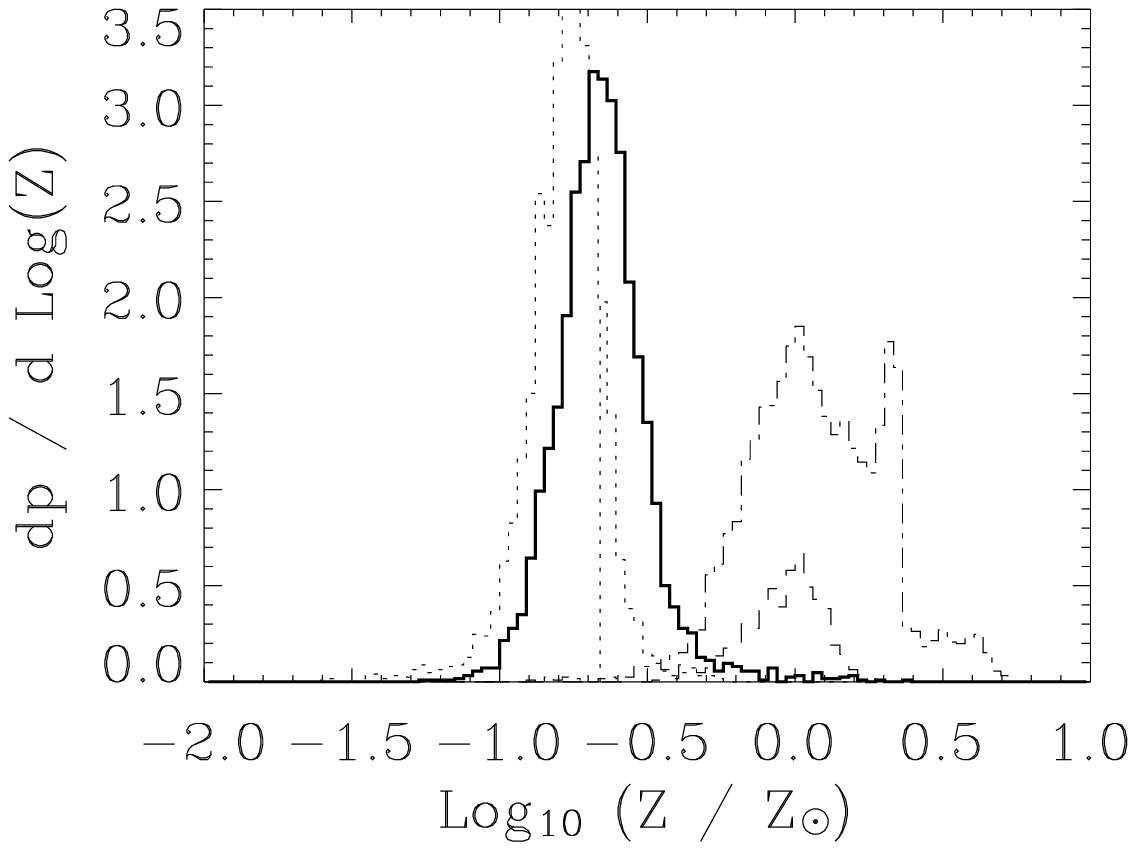,width=\hssize} & 
\psfig{figure=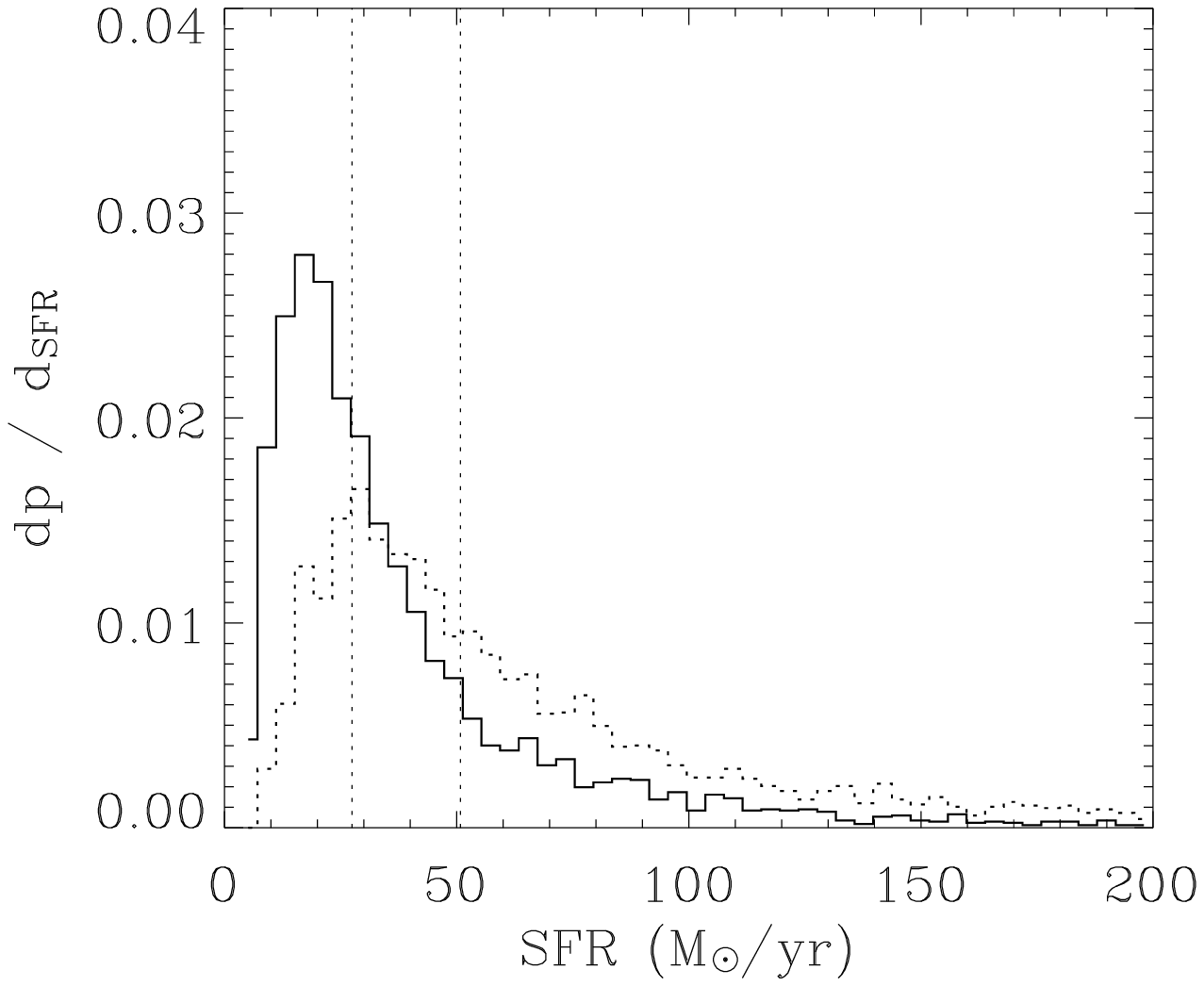,width=\hssize} \\
\end{tabular}
\end{center}
\caption{{\it Top left panel :} Radius (kpc) distribution for our
modelled LBGs.  The radius is a stellar-mass weighted average radius
of three components: disc, bulge and starburst.  The median value of $\sim
2.6$ kpc, indicated with the dotted vertical line, is in good agreement
with the data from the HDF \citep{GiavaliscoSteidelMacchetto96}.  {\it Top right
panel :} Stellar mass distributions for our modelled LBGs. The median
value of $2.6\times10^{10}$M$_{\odot}$ is indicated by the dotted
vertical line.  {\it Bottom left panel :} The solid histogram shows the 3-component average ISM
metallicity distribution of our modelled galaxies, the dotted histogram that of the
discs, the dashed histogram that of the bulges and
the dot-dashed one that of the starbursts.  {\it Bottom right
panel :} The solid histograms represent the distribution of instantaneous SFRs of our modelled LBGs, whereas the dotted histogram 
shows the distribution of SFRs averaged over 100 Myr for the same galaxies. The
median SFR$_{inst}$ is $\sim$28 M$_{\odot}$ yr$^{-1}$, as indicated by the
left hand side vertical dotted line, and the median of SFR$_{100}$ is $\sim$50.8 M$_{\odot}$ yr$^{-1}$, indicated by the 
right hand side vertical dotted line.}
\label{fig:LBG_props}	
\end{figure*}

\subsubsection*{Sizes}
In Fig. \ref{fig:LBG_props} we show the distribution of sizes of our
modelled LBGs. The size of a galaxy is defined here as the
stellar-mass weighted average of the three component (disc, bulge, starburst) half-mass
radii. We find a median value of $2.6$ kpc for our LBGs, which is in
good agreement with the analysis by
\citet{GiavaliscoSteidelMacchetto96} of a few LBGs observed with the
HST (but identified with ground-based images) for which they find a
typical size of about $2$ kpc.  The mean values and standard
deviations are summarised in table \ref{tab:general_props}.

\subsubsection*{Stellar masses} \label{sec:stellar_mass}
The distribution of stellar masses of our modelled LBGs is given in
Fig. \ref{fig:LBG_props}. The median stellar mass is $2.6\times 10^{10}$
M$_{\odot}$, in very good agreement with the mean stellar mass deduced
from data by \citet{ShapleyEtal01} ($<M_{star}>_{med}=2.4\times 10^{10}$
M$_{\odot}$), and \citet{PapovichDickinsonFerguson01} ($\sim 10^{10}$
M$_{\odot}$). 
The shape of our distribution is consistent with that given by
Shapley et al. (2001, their Fig. 10b), and also with the ``accelerated
quiescent'' model of SPF \citep{PrimackWechslerSomerville01}.

\subsubsection*{Metallicity}
The distribution of metallicities of our modelled LBGs is given in
Fig. \ref{fig:LBG_props}. The solid histogram shows the 3-component average ISM
metallicity of our modelled galaxies, the dotted histogram that of the
discs, the dashed line shows that of the bulges and
the dot-dashed one that of the starbursts. We find a median
metallicity of about $Z_{\odot}/5$, which is again in good agreement
with observations \citep{PettiniEtal01}. The starburst and bulge components have higher
metallicities, due to the more intensive star formation taking place there.

\subsubsection*{Star formation rates}
In Fig. \ref{fig:LBG_props} we show the distribution of star formation
rates (SFR) for our modelled LBGs computed in two different ways.  The
solid line shows the instantaneous star formation rate (SFR$_{inst}$),
computed over the last Myr. This quantity is not directly comparable
to the value deduced from observations of the UV through broad band
filters (see discussion in \citet{Kennicutt98a}). However, they
indicate the current state of LBGs, and should compare better to
estimates of SFR obtained from emission lines. For this estimate, we
find a median value of 27.4 M$_{\odot}$ yr$^{-1}$, lower than the
observed value by a factor $\sim 3$.  The dotted curve shows the
distribution of star formation rates computed as the mass of stars
younger than 100 Myr, divided by 100 Myr (SFR$_{100}$). This is much
closer to what is deduced from observations by \citet{ShapleyEtal01}
as it takes into account the star formation history during a time of
order the lifetime of B and A stars. Our measure is however an
under-estimate of the real SFR (by about 20\%) because we miss all the
mass lost through stellar evolution during a 100 Myr. The median
value of SFR$_{100}$ is 50.8 M$_{\odot}$ yr$^{-1}$, with 20\% of the modelled LBGs 
forming more that 100  M$_{\odot}$ per year. These SFRs are in better agreement with the
data, although still low compared to the estimate of
\citet{ShapleyEtal01} : $\sim 90$ M$_{\odot}$ yr$^{-1}$. 

The question of what triggers these high SFRs has been investigated by \citet{SomervillePrimackFaber01} 
who suggest that star formation in LBGs is triggered by mergers. In our model, we find that 
only 30\% of the LBGs have undergone a merger (although this figure goes up to 60\% for
our high resolution simulation). Thus we conclude that star formation in LBGs is both due 
to merger-triggered starbursts and active cooling of gas on the galaxies.

\subsubsection*{Conclusion}
The main point that the previous statistics make is that \gal\ does not 
reproduce LBG counts by chance, as 
it reproduces their dynamical and chemical properties quite well. Again, 
we emphasise that this is a natural outcome of a 
model which also reproduces quite well the properties of local galaxies, 
as shown in \hatton.

\begin{table*} 
\begin{center}
\begin{minipage}{15.5cm} \def\footnoterule{}
\begin{center}
\begin{tabular}{lccccc} 
\hline\hline
LBGs        & $<r_{1/2}>$ \footnote{Half-mass radius of largest galaxy component (disc, bulge, or burst) (standard deviation).}  &    $<M_{star}>$   \footnote{Stellar mass of galaxies (standard deviation).}  &      $<Z>$  \footnote{Mean metallicity (standard deviation).} & $<$SFR$_{inst}>$\footnote{Instantaneous star formation rate summed over the three components of our modelled galaxies.}& $<$SFR$_{100}>$\footnote{Star formation rate summed over the three components of our modelled galaxies, and averaged over 100 Myr.} \\
            & (kpc) &($10^{10} M_{\odot}$)& ($Z_{\odot}$) &(M$_{\odot}$ yr$^{-1}$) & (M$_{\odot}$ yr$^{-1}$)\\ \hline
$R\leq25.5$ & 2.61  &  2.60               & 0.219         & 27.4                 &50.8 \\
$R\leq25$   & 2.61  &  2.52               & 0.221         & 26.2                 &48.4 \\
$R\leq24.5$ & 2.59  &  2.20               & 0.217         & 23.8                 &43.2 \\ \hline
\end{tabular}
\label{tab:general_props}
\end{center}
\end{minipage}
\caption{General properties of our modelled LBGs at \zz.}
\end{center}
\end{table*}

\subsection{The descendants of LBGs} \label{sec:descendants}
In a recent paper, \citet{Nagamine02} investigated the nature of the progenitors and
descendants of LBGs within a cosmological hydrodynamical simulation. We extend here
this investigation to the new methodology we have developed with \gal{} and \momaf{}.
We first start with examples of star formation histories of modelled LBGs, following them until 
$z=0$, and continue with an investigation of the nature of the local descendants of these galaxies.

\subsubsection{Star formation histories} \label{sec:SFHs}
\begin{figure*}
\centerline{\psfig{figure=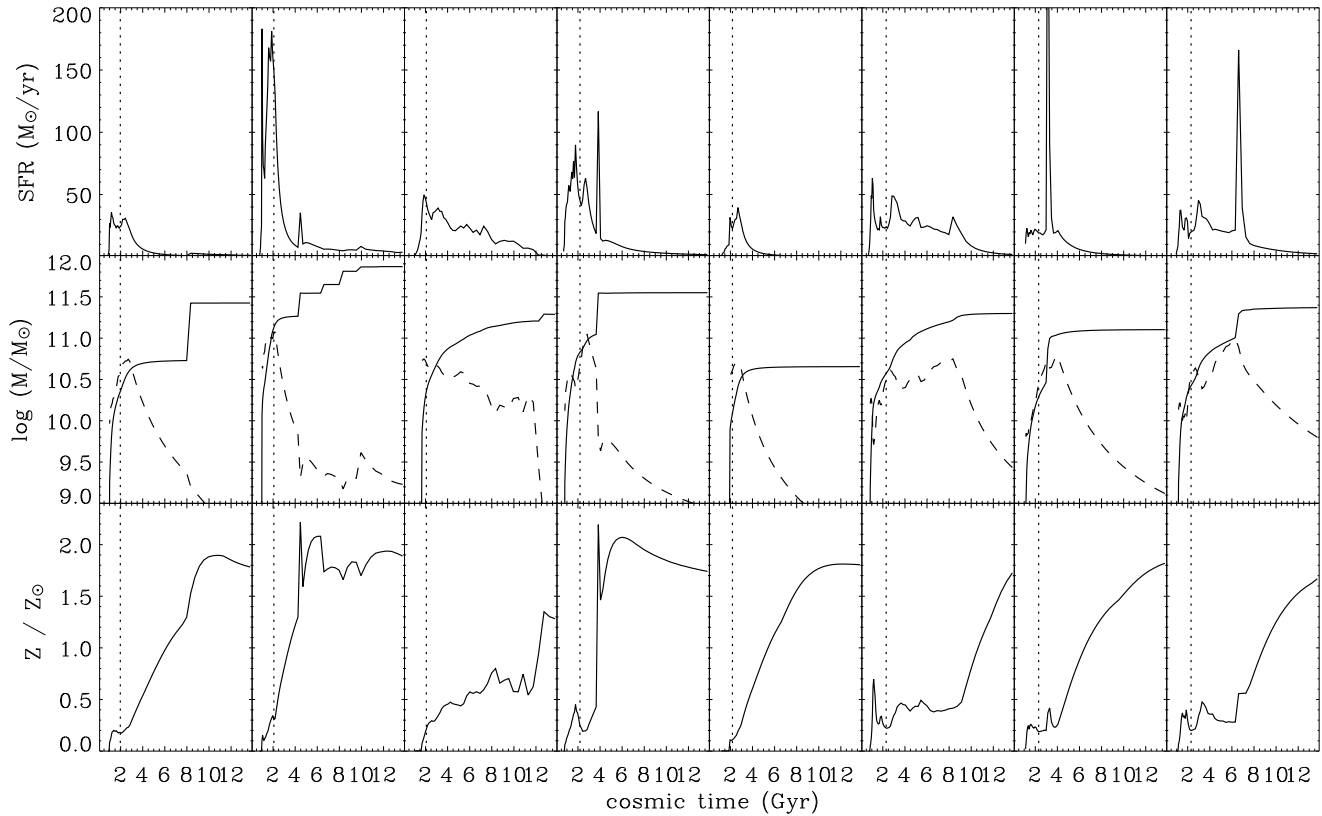,width=\hdsize}}
\caption{Examples of galaxy histories in a $\Lambda$CDM cosmology. {\it Top panels} : instantaneous SFR versus
cosmic time. {\it Middle panel} : masses of stars (solid curves) and
gas (dashed curves) in our LBGs versus cosmic
time. {\it Lower panels} : evolution of average ISM metallicities with cosmic
time. The vertical dotted lines indicate the time at which galaxies were
identified as LBGs.}
\label{fig:history}
\end{figure*}

In Fig. \ref{fig:history} we show several examples of star formation
histories for our modelled LBGs. The top panels show the star formation
rate as a function of cosmic time; the middle panels the
stellar (solid curves) and gas (dashed curve) mass evolution; 
and the lower panels that of the metallicity.
The vertical dotted lines indicate the moment at which 
each galaxy is identified as a LBG. Before this vertical line, the given 
history is that of the most massive progenitor at each merger undergone by the future LBG.
Each column represents a galaxy, and they are sorted from
left to right in order of decreasing redshift of identification.

We find a large variety of star formation histories, representative of the
variety of states in which galaxies can be when they meet the LBG criteria.
This also illustrates the two processes that trigger star formation in LBGs : active cooling
of the hot gas in halos onto the young galaxies as opposed to merger-triggered starbursts.

Finally, notice how some LBGs evolve quiescently until the present
day, resulting in massive spirals of which they form the bulges, and
how some undergo several extra mergers, resulting in present-day
massive ellipticals. From the middle panels of Fig. \ref{fig:history}
it can be seen that our LBGs represent, on average, about 10\% of the
total mass of their descendants in the local Universe.

\begin{figure}
\centerline{\psfig{figure=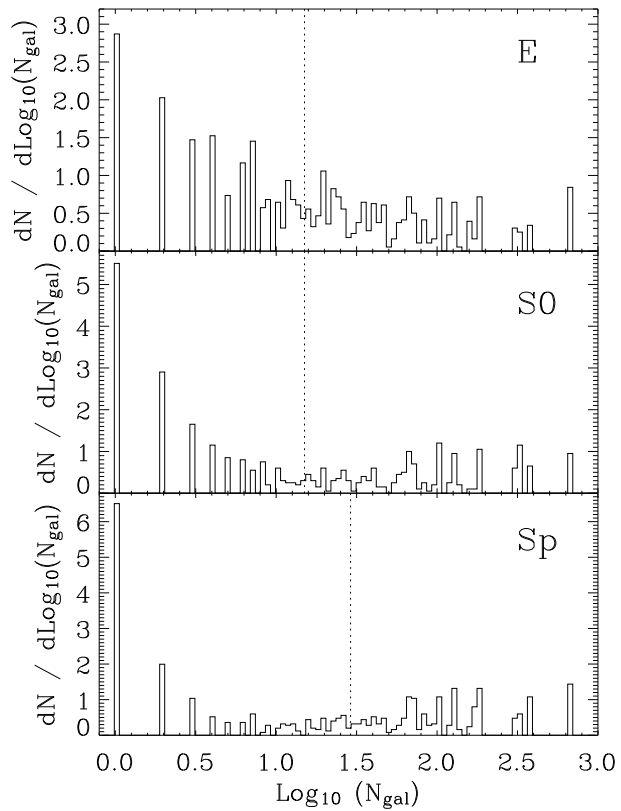,width=\hssize}}
\caption{Number of galaxies in haloes where LBG descendants are found, 
for three morphological types of descendants. Vertical dotted lines indicate median numbers.}
\label{fig:cumul_N_in_groups}
\end{figure}

\subsubsection{LBGs as progenitors of ellipticals ?}
We define the morphology of galaxies at $z=0$ using the ratio of bulge to 
disc B-band luminosities, as described in \hatton. With this definition, 
we find that $\sim$77\% of our modelled LBGs end up in ellipticals or 
lenticular galaxies, the remainder ending up in spirals. On the other hand,
we also computed the number of elliptical or lenticular galaxies at $z=0$ that 
have had at least one LBG progenitor in their merging history tree. These galaxies 
represent about 35\% of the whole local population of early-type galaxies (with $M_B<-18$).

In Fig. \ref{fig:cumul_N_in_groups} we show the expected location of the 
descendants of our modelled LBGs. The x-axis is the size (i.e. number 
of galaxies) of groups or clusters of galaxies, and the y-axis gives the 
number of LBGs that end up in groups of these sizes. Elliptical or 
lenticular descendants of LBGs are found in groups with a median number of 
$\sim$15 galaxies, while spiral descendants are found to be satellite 
galaxies in clusters/groups of $\sim$30 galaxies.


\section{Discussion} \label{sec:conclusion}

\subsection{Summary of results}
This paper is the third one of the \gal~series that presents a consistent 
attempt to model hierarchical galaxy formation within a hybrid approach in 
which dark matter collapse is computed in a large cosmological simulation, 
and the physics of baryons is described by semi--analytic recipes. 
The first paper (\hatton) has detailed the model and the predicted 
statistical properties of local galaxies, which have been compared against 
data. The second paper (\devriendta) has presented the evolution of a set of 
these properties from redshift $z=3$ to $z=0$ within the hierarchical 
paradigm. In the current paper, we have proposed a more detailed study of the 
properties of optically--bright, star--forming galaxies at $z \sim 3$. 
We have converted the predictions of our \gal~model
into a mock observing cone by means of 
the \mo~package. From this observing cone, a sample of LBGs
has been extracted with selection criteria as close as possible to those 
used in the observational sample of \citet{SteidelEtal96}. This method 
allows us to get a three--fold insight on (i) the efficiency of the 
selection criteria to capture $z \sim 3$ galaxies, (ii) 
the nature of the objects that have been selected according to
these criteria, and (iii) the properties that could be measured in forthcoming 
follow--up. The results of this paper have been obtained from the same 
cosmological simulation (a $\Lambda$CDM model with $256^3$ particles in a 
cubic box with $100h^{-1}$ Mpc on a side), and with the same astrophysical 
parameters in the semi--analytic post--processing as in the first 
two papers. In doing so, we are attempting to progressively build up 
a consistent view of hierarchical galaxy formation from $z=3$ to $z=0$.

In this prospect, the successes of the model are satisfactory, in spite of
several shortcomings. Our model LBGs were selected 
according to colour and magnitude criteria in the mock observing cone. This 
allowed us first to examine the efficiency of the selection method. 
We have seen that:

\begin{itemize}
  
  \item The model is able to reproduce the number density counts of LBGs 
  on the sky, and gives a number density of $\sim$1.2 arcmin$^{-2}$ at 
  the magnitude limit of the sample, in good agreement with the data.

  \item The set of colour and magnitude criteria designed by
  \citet{SteidelHamilton93} and \citet{SteidelPettiniHamilton95} to
  select $z \sim 3 $ galaxies on the basis of synthetic stellar
  populations evolving according to the monolithic model, is well
  suited to the selection of LBGs within the hierarchical model. Our
  best contour in the colour--colour plot is not very different from the
  usual contour.

  \item The fraction of interlopers, that is, of galaxies that pass the 
  criteria, but are not located in the target redshift range $[2.7,3.4]$, 
  is about 12\%. Most of these interlopers are foreground galaxies at $z>2$.  

  \item The efficiency of the selection, defined as the product of the 
  {\em completeness} by the {\em confirmation rate}, is 85\%.

  \item The contribution of clustering to cosmic variance can be estimated 
  for scales below the apparent size of the box at redshift $\sim 3$. For a 
  100 arcmin$^2$ field, the relative variance is more than twice the 
  Poissonian value.

  \item The reconstructed luminosity function, provided we apply a
  correction for volume effects as in
  \citet{SteidelPettiniHamilton95}, is very similar to the one
  given in that paper. This shows an overall consistency between
  the mock sample and the actual sample.

\end{itemize}
 
Given this overall agreement, we have explored the nature of LBGs in the model,
with special emphasis on three critical issues (i) clustering and
spatial information, (ii) extinction and the optical--to--IR luminosity 
budget, and (iii) merging history and fate in the present universe. We have obtained 
the following results: 

\begin{itemize}

   \item The amplitude and shape of the projected 2--point correlation function 
   is in agreement with the data. With our choice of the 
   cosmological parameters of the simulation, the comparison of the 3D 
   correlation function with the dark matter correlation function gives us an 
   estimate of the linear bias that slowly decreases on scales from 1 to 20 
   $h^{-1}$Mpc, from $b \simeq 5$ to $b \simeq 3$. So we find a higher bias value 
   than previously found in the literature. This is the first estimate of 
   the bias for $z \sim 3$ LBGs from a hybrid model. The LBGs are located 
   in haloes with masses ranging from $3 \times 10^{11} M_\odot$ to
   $3 \times 10^{12} M_\odot$, 
   with a median value  $1.6 \times 10^{11} M_\odot$. The positions of the LBGs 
   nicely delineate the dense regions of the filaments at $z \sim 3$. 

   \item The extinction properties of our model LBGs depend on the
   extinction model that has been put in, basically the one presented
   in the {\sc stardust} model of spectrophotometric evolution
   \citep{DevriendtGuiderdoniSadat99}. We find a factor $\sim 6.2$ for
   extinction at the rest--frame wavelength 1600 \AA\, that is, a
   value higher than that used by \citet{SteidelEtal96}, but
   consistent with the later analysis of
   \citet{AdelbergerSteidel00}. The UV to IR luminosity budget is
   quite similar to the re--construction proposed by
   \citet{AdelbergerSteidel00} on the basis of the available data.

   \item We estimated a set of properties for our mock LBGs. The
   median values are $2.6 \times 10^{10} M_\odot$ for stellar masses,
   $2.6$ kpc for half--light radii, $0.2 \ Z_\odot$ for
   metallicities, and 27 (resp. 50) M$_\odot$yr$^{-1}$ for instantaneous (resp. averaged over 100 Myr) SFRs. These median values
   for stellar masses, radii and metallicities are in agreement with
   the data, although the measurement of these quantities from
   observations is not straightforward.  Our value for the median SFR
   seems to be too low with respect to observations\footnote{We believe, however, that this underestimate is due
   both to differences in stellar population models used to retrieve the SFR and in the
   definition of the period on which the SFR is computed. When playing
   the game of fitting our LBG colours with e.g. the PEGASE model
   \citep{FiocRocca97}, we obtain larger values of the SFR, by a
   factor 2--3.}. We also find that only a minority of LBGs (30\%) may be
   starbursts triggered by mergers, contrary to \citet{SomervillePrimackFaber01}.

   \item Finally, we are able to assess the future of these $z \sim 3$ LBGs. 
   We give examples of their SFR and mass histories from this epoch to the 
   present. We are also interested in the nature of LBGs as progenitors of 
   current galaxies of the Hubble sequence, although morphological 
   classification issues are not easy. We recall that in \gal, our Hubble 
   type is given by the $B$--band bulge--to--disk luminosity ratio. 
   The model LBGs observed at 
   $z \sim 3$ preferentially end up in galaxies that are classified as 
   bright, early--type galaxies. More specifically, 77 \% of the descendants 
   of our model LBGs are ellipticals or lenticulars with $M_B<-18$,
   whereas the latter population only represents 15 \% of the local galaxies 
   with $M_B<-18$. On the other hand, 35 \% of the bright ellipticals and 
   lenticulars of the local Universe actually had a LBG progenitor at 
   $z \sim 3$. The other 65 \% may have had already progenitors at 
   $z \sim 3$, but with apparent magnitudes that are to faint to pass 
   the chosen selection threshold. So we cannot say that LBGs ``\`a la 
   Steidel'' are {\em the} progenitors of local ellipticals, but they clearly 
   are progenitors of local ellipticals.

\end{itemize}

Our model gives a very good basis for making predictions of 
observational strategies for forthcoming follow--up observations. 
For instance, we show that only a 
small fraction of the LBGs can be detected in a SCUBA follow-up or with the 
SWIRE\footnote{{\tt http://www.ipac.caltech.edu/SWIRE/}} survey with SIRTF. 
These objects are 
clearly a key target for ALMA which should be able to detect all of them at 
850 $\mu$m or $1.3$ mm, at the 0.1 mJy level. The strong potential of our 
model for making follow--up predictions through many optical and IR/submm
filters can be fully exploited by submitting 
queries to our relational database at {\tt http://galics.iap.fr}. 

\subsection{Summary of drawbacks}
The main shortcoming of our model for the present study comes from the mass 
resolution of the cosmological simulation we use. Only haloes with 20 or 
more particles are identified by the {\sc fof} algorithm. This results 
in a baryonic mass limit, below which incompleteness comes in. 
Galaxies with baryonic masses below the threshold are included in identified 
haloes, in which not all the hot gas has cooled, whereas galaxies in 
dark matter structures less massive than our detection threshold are 
missing. Fortunately, we have shown in \hatton~that incompleteness settles 
slowly for decreasing masses. On its turn, the baryonic mass threshold 
converts into an absolute magnitude limit that we call our formal magnitude 
limit. The estimate of this value at $z \sim 3$ for rest--frame 1600 \AA\ 
absolute magnitude is $M_{AB}=-20.3$. This means that this mass resolution 
just passes the  selection criterion put as $M_{AB} \leq - 20.2$. 

Unfortunately, the mass resolution was shown to ``leak'' on galaxy 
properties through their merging histories by \hatton. It was not crucial 
for bright galaxies at $z=0$, but it is much more significant at $z \sim 
3$. Galaxies that should have been missed because they are not massive enough
appear over the magnitude threshold 
because they are ``immature''. Their history is not resolved enough in the 
past, because of the absence of small haloes, and they appear 
too bright inasmuch as they are too gassy and form stars too actively. 
We showed that we can recover from this 
effect by imposing a ``maturity criterion'': only galaxies with at least
a progenitor 1.1 Gyr ago have to be selected. The other objects have to be discarded
from our analysis. Such a criterion admittedly appears as a fine--tuning to recover 
the correct number of LBGs. It is only at $z=3$ that we are 'drowning' in major resolution effects, 
inasmuch as the correction reaches about 75 \%, 
and we are consequently reaching the limit of this specific simulation.
A study on a higher--resolution simulation (Blaizot et al, 2004, in progress) 
shows that the chosen maturity criterion becomes 
transparent at $z=3$ in this simulation, and that our maturity criterion to by--pass 
the problem in the current simulation is reasonable.

Another shortcoming appears as the consequence of the 
volume of our cosmological simulation. The comoving side of the box 
100$h^{-1}$ Mpc corresponds to an apparent 1 deg at $z \sim 3$. It
does not allow us to address 
the issue of cosmic variance in fields larger than typically a fraction of a 
deg$^2$, and clustering on scales larger than a fraction of 
a degree. On scales smaller than that, we have shown in a companion paper 
(\mo) that the technique of cone building from this simulation introduces 
systematic effects that lower the correlation function at most by 10 \%.

Finally, we have tried to model the selection criteria as accurately as 
possible, but we may have missed systematic effects and subtle observational 
biases. 

\section{Conclusions}
The LBG selection is a very efficient way to select high--redshift galaxies.
The $z \sim 3$ LBGs seem to naturally come out of models of hierarchical 
galaxy formation, as shown by the current study. The list of properties 
that are reproduced is quite impressive, and the overall feeling is that we 
globally understand what is going on, both in the selection process and 
in the nature of the target galaxies. However, the actual impact of this 
set of results is somewhat affected by the need to correct for the 
``immaturity'' of a large number of $z \sim 3$ galaxies in the model. This
immaturity reflects mass resolution in terms of history resolution. 
For this reason, there is no point in attempting to go at higher 
redshift, and try to model 
$z \sim 4$ LBGs \citep{SteidelEtal99} with the outputs of this specific simulation.

The \gal~model seems to nicely mimic the properties of 
optically--bright, star--forming galaxies at $z \sim 3$. However, this 
obviously does not mean that other galaxy populations in this redshift 
range that would be selected according to other observational criteria, 
are equally well reproduced. More specifically, samples of galaxies may be 
selected for instance on the basis of their red colours, such as the 
{\em Extremely Red Objects}. In this prospect, another subtle effect of mass resolution may 
appear in terms of the absence of old stellar populations that should 
have formed in structures smaller than the mass threshold. This issue will be 
addressed in forthcoming work.

After our first two \gal~papers that presented a detailed exploration of the 
nature of model galaxies between $z=0$ and 3, this study is the next
milestone in the presentation of the full sets of results 
coming out of the \gal~model, as far as it illustrates how mock samples can 
be extracted with elaborated selection criteria. This study is complemented 
by a relational database that can be easily queried through a web interface 
at {\tt http://galics.iap.fr}.
Two forthcoming papers will
be more oriented toward predictions for deep surveys, in terms of 
multi--wavelength faint galaxy counts (\devriendtb) and 2D correlation 
functions (\blaizotb).

The road map for further developments is rather clear. On the one hand, 
since we are reaching the limit of this specific simulation, we have to run 
simulations with a better resolution. Such a work is in progress and seems 
to nicely confirm the results of the current study. On the other hand, the 
logic of mock observing cones should lead us to generate mock images with all 
the instrumental effects, and to try to extract mock samples with the 
same data processing pipelines as the actual observations. 


\section*{Acknowledgments}
We thank S. Colombi for his precious help on correlation 
functions and error estimates, and D. Le Borgne for helping us with the PEGASE web interface. 
JEGD and SH were respectively supported by  
grants from the Leverhulme trust and the EU TMR network {\em Formation and 
Evolution of Galaxies}. FS acknowledges support from the Marie Curie EARASTARGAL program.


\begin{thebibliography}{}

\bibitem[\protect\citeauthoryear{{Adelberger} \& {Steidel}}{{Adelberger} \&
  {Steidel}}{2000}]{AdelbergerSteidel00}
{Adelberger} K.~L.,  {Steidel} C.~C.,  2000, \apj, 544, 218

\bibitem[\protect\citeauthoryear{{Adelberger}, {Steidel}, {Giavalisco},
  {Dickinson}, {Pettini} \& {Kellogg}}{{Adelberger}
  et~al.}{1998}]{AdelbergerEtal98}
{Adelberger} K.~L.,  {Steidel} C.~C.,  {Giavalisco} M.,  {Dickinson} M.,
  {Pettini} M.,    {Kellogg} M.,  1998, \apj, 505, 18

\bibitem[\protect\citeauthoryear{{Arnouts}, {Moscardini}, {Vanzella},
  {Colombi}, {Cristiani}, {Fontana}, {Giallongo}, {Matarrese} \&
  {Saracco}}{{Arnouts} et~al.}{2002}]{ArnoutsEtal02}
{Arnouts} S.,  {Moscardini} L.,  {Vanzella} E.,  {Colombi} S.,  {Cristiani} S.,
   {Fontana} A.,  {Giallongo} E.,  {Matarrese} S.,    {Saracco} P.,  2002,
  \mnras, 329, 355

\bibitem[\protect\citeauthoryear{{Balland}, {Devriendt} \& {Silk}}{{Balland}
  et~al.}{2003}]{BallandDevriendtSilk03}
{Balland} C.,  {Devriendt} J.~E.~G.,    {Silk} J.,  2003, \mnras, 343, 107

\bibitem[\protect\citeauthoryear{{Baugh}, {Cole}, {Frenk} \& {Lacey}}{{Baugh}
  et~al.}{1998}]{BaughEtal98}
{Baugh} C.~M.,  {Cole} S.,  {Frenk} C.~S.,    {Lacey} C.~G.,  1998, \apj, 498,
  504

\bibitem[\protect\citeauthoryear{{Blaizot}, {Wadadekar}, {Guiderdoni},
  {Colombi}, {Bertin}, {Bouchet}, {Devriendt} \& {Hatton}}{{Blaizot}
  et~al.}{2003}]{BlaizotEtal03a}
{Blaizot} J.,  {Wadadekar} Y.,  {Guiderdoni} B.,  {Colombi} S.,  {Bertin} E.,
  {Bouchet} F.~R.,  {Devriendt} J.~E.~G.,    {Hatton} S.,  2003, submitted to
  MNRAS

\bibitem[\protect\citeauthoryear{{Chapman}, {Scott}, {Steidel}, {Borys},
  {Halpern}, {Morris}, {Adelberger}, {Dickinson}, {Giavalisco} \&
  {Pettini}}{{Chapman} et~al.}{2000}]{ChapmanEtal00}
{Chapman} S.~C.,  {Scott} D.,  {Steidel} C.~C.,  {Borys} C.,  {Halpern} M.,
  {Morris} S.~L.,  {Adelberger} K.~L.,  {Dickinson} M.,  {Giavalisco} M.,
  {Pettini} M.,  2000, \mnras, 319, 318

\bibitem[\protect\citeauthoryear{{Devriendt} \& {Guiderdoni}}{{Devriendt} \&
  {Guiderdoni}}{2000}]{DevriendtGuiderdoni00}
{Devriendt} J.~E.~G.,  {Guiderdoni} B.,  2000, \aap, 363, 851

\bibitem[\protect\citeauthoryear{{Devriendt}, {Guiderdoni} \&
  {Sadat}}{{Devriendt} et~al.}{1999}]{DevriendtGuiderdoniSadat99}
{Devriendt} J.~E.~G.,  {Guiderdoni} B.,    {Sadat} R.,  1999, \aap, 350, 381

\bibitem[\protect\citeauthoryear{{Devriendt}, {Hatton}, {Ninin}, {Blaizot},
  {Bouchet} \& {Guiderdoni}}{{Devriendt} et~al.}{2003}]{DevriendtEtal03}
{Devriendt} J.~E.~G.,  {Hatton} S.,  {Ninin} S.,  {Blaizot} J.,  {Bouchet}
  F.~R.,    {Guiderdoni} B.,  2003, in preparation

\bibitem[\protect\citeauthoryear{{Fioc} \& {Rocca-Volmerange}}{{Fioc} \&
  {Rocca-Volmerange}}{1997}]{FiocRocca97}
{Fioc} M.,  {Rocca-Volmerange} B.,  1997, \aap, 326, 950

\bibitem[\protect\citeauthoryear{{Giavalisco} \& {Dickinson}}{{Giavalisco} \&
  {Dickinson}}{2001}]{GiavaliscoDickinson01}
{Giavalisco} M.,  {Dickinson} M.,  2001, \apj, 550, 177

\bibitem[\protect\citeauthoryear{{Giavalisco}, {Steidel}, {Adelberger},
  {Dickinson}, {Pettini} \& {Kellogg}}{{Giavalisco}
  et~al.}{1998}]{GiavaliscoEtal98}
{Giavalisco} M.,  {Steidel} C.~C.,  {Adelberger} K.~L.,  {Dickinson} M.~E.,
  {Pettini} M.,    {Kellogg} M.,  1998, \apj, 503, 543

\bibitem[\protect\citeauthoryear{{Giavalisco}, {Steidel} \&
  {Macchetto}}{{Giavalisco} et~al.}{1996}]{GiavaliscoSteidelMacchetto96}
{Giavalisco} M.,  {Steidel} C.~C.,    {Macchetto} F.~D.,  1996, \apj, 470, 189

\bibitem[\protect\citeauthoryear{{Guiderdoni}, {Hivon}, {Bouchet} \&
  {Maffei}}{{Guiderdoni} et~al.}{1998}]{GuiderdoniEtal98}
{Guiderdoni} B.,  {Hivon} E.,  {Bouchet} F.~R.,    {Maffei} B.,  1998, \mnras,
  295, 877

\bibitem[\protect\citeauthoryear{{Hatton}, {Devriendt}, {Ninin}, {Bouchet},
  {Guiderdoni} \& {Vibert}}{{Hatton} et~al.}{2003}]{HattonEtal03}
{Hatton} S.,  {Devriendt} J.~E.~G.,  {Ninin} S.,  {Bouchet} F.~R.,
  {Guiderdoni} B.,    {Vibert} D.,  2003, \mnras, 343, 75

\bibitem[\protect\citeauthoryear{{Idzi}, {Somerville}, {Papovich}, {Ferguson},
  {Giavalisco}, {Kretchmer} \& {Lotz}}{{Idzi} et~al.}{2003}]{IdziEtal03}
{Idzi} R.,  {Somerville} R.,  {Papovich} C.,  {Ferguson} H.~C.,  {Giavalisco}
  M.,  {Kretchmer} C.,    {Lotz} J.,  2003, ArXiv Astrophysics e-prints

\bibitem[\protect\citeauthoryear{{Kennicutt}}{{Kennicutt}}{1983}]{Kennicutt83}
{Kennicutt} R.~C.,  1983, \apj, 272, 54

\bibitem[\protect\citeauthoryear{{Kennicutt}}{{Kennicutt}}{1998}]{Kennicutt98a}
{Kennicutt} R.~C.,  1998, \araa, 36, 189

\bibitem[\protect\citeauthoryear{{Landy} \& {Szalay}}{{Landy} \&
  {Szalay}}{1993}]{LandySzalay93}
{Landy} S.~D.,  {Szalay} A.~S.,  1993, \apj, 412, 64

\bibitem[\protect\citeauthoryear{{Madau}}{{Madau}}{1995}]{Madau95}
{Madau} P.,  1995, \apj, 441, 18

\bibitem[\protect\citeauthoryear{{Madau}, {Ferguson}, {Dickinson},
  {Giavalisco}, {Steidel} \& {Fruchter}}{{Madau} et~al.}{1996}]{MadauEtal96}
{Madau} P.,  {Ferguson} H.~C.,  {Dickinson} M.~E.,  {Giavalisco} M.,  {Steidel}
  C.~C.,    {Fruchter} A.,  1996, \mnras, 283, 1388

\bibitem[\protect\citeauthoryear{{Massarotti}, {Iovino} \&
  {Buzzoni}}{{Massarotti} et~al.}{2001}]{MassarottiIovinoBuzzoni01}
{Massarotti} M.,  {Iovino} A.,    {Buzzoni} A.,  2001, \apjl, 559, L105

\bibitem[\protect\citeauthoryear{{Nagamine}}{{Nagamine}}{2002}]{Nagamine02}
{Nagamine} K.,  2002, \apj, 564, 73

\bibitem[\protect\citeauthoryear{{Ninin}}{{Ninin}}{1999}]{Ninin99}
{Ninin} S.,  1999, Th\`ese de l'Universit\'e Paris 11

\bibitem[\protect\citeauthoryear{{Papovich}, {Dickinson} \&
  {Ferguson}}{{Papovich} et~al.}{2001}]{PapovichDickinsonFerguson01}
{Papovich} C.,  {Dickinson} M.,    {Ferguson} H.~C.,  2001, \apj, 559, 620

\bibitem[\protect\citeauthoryear{{Pettini}, {Shapley}, {Steidel}, {Cuby},
  {Dickinson}, {Moorwood}, {Adelberger} \& {Giavalisco}}{{Pettini}
  et~al.}{2001}]{PettiniEtal01}
{Pettini} M.,  {Shapley} A.~E.,  {Steidel} C.~C.,  {Cuby} J.,  {Dickinson} M.,
  {Moorwood} A.~F.~M.,  {Adelberger} K.~L.,    {Giavalisco} M.,  2001, \apj,
  554, 981

\bibitem[\protect\citeauthoryear{{Porciani} \& {Giavalisco}}{{Porciani} \&
  {Giavalisco}}{2002}]{PorcianiGiavalisco02}
{Porciani} C.,  {Giavalisco} M.,  2002, \apj, 565, 24

\bibitem[\protect\citeauthoryear{{Press}, {Teukolsky}, {Vetterling} \&
  {Flannery}}{{Press} et~al.}{1992}]{NumericalRecipes}
{Press} W.~H.,  {Teukolsky} S.~A.,  {Vetterling} W.~T.,    {Flannery} B.~P.,
  1992, {Numerical recipes in FORTRAN. The art of scientific computing}.
Cambridge: University Press, |c1992, 2nd ed.

\bibitem[\protect\citeauthoryear{{Primack}, {Wechsler} \&
  {Somerville}}{{Primack} et~al.}{2001}]{PrimackWechslerSomerville01}
{Primack} J.~R.,  {Wechsler} R.~H.,    {Somerville} R.~S.,  2001, in {Bender}
  R.,  {Renzini} A.,  eds, Mass of Galaxies at Low and High Redshift The masses
  of lyman break galaxies

\bibitem[\protect\citeauthoryear{{Shapley}, {Steidel}, {Adelberger},
  {Dickinson}, {Giavalisco} \& {Pettini}}{{Shapley}
  et~al.}{2001}]{ShapleyEtal01}
{Shapley} A.~E.,  {Steidel} C.~C.,  {Adelberger} K.~L.,  {Dickinson} M.,
  {Giavalisco} M.,    {Pettini} M.,  2001, \apj, 562, 95

\bibitem[\protect\citeauthoryear{{Somerville}, {Primack} \&
  {Faber}}{{Somerville} et~al.}{2001}]{SomervillePrimackFaber01}
{Somerville} R.~S.,  {Primack} J.~R.,    {Faber} S.~M.,  2001, \mnras, 320, 504

\bibitem[\protect\citeauthoryear{{Steidel}, {Adelberger}, {Giavalisco},
  {Dickinson} \& {Pettini}}{{Steidel} et~al.}{1999}]{SteidelEtal99}
{Steidel} C.~C.,  {Adelberger} K.~L.,  {Giavalisco} M.,  {Dickinson} M.,
  {Pettini} M.,  1999, \apj, 519, 1

\bibitem[\protect\citeauthoryear{{Steidel}, {Giavalisco}, {Pettini},
  {Dickinson} \& {Adelberger}}{{Steidel} et~al.}{1996}]{SteidelEtal96}
{Steidel} C.~C.,  {Giavalisco} M.,  {Pettini} M.,  {Dickinson} M.,
  {Adelberger} K.~L.,  1996, \apjl, 462, L17+

\bibitem[\protect\citeauthoryear{{Steidel} \& {Hamilton}}{{Steidel} \&
  {Hamilton}}{1993}]{SteidelHamilton93}
{Steidel} C.~C.,  {Hamilton} D.,  1993, \aj, 105, 2017

\bibitem[\protect\citeauthoryear{{Steidel}, {Pettini} \& {Hamilton}}{{Steidel}
  et~al.}{1995}]{SteidelPettiniHamilton95}
{Steidel} C.~C.,  {Pettini} M.,    {Hamilton} D.,  1995, \aj, 110, 2519

\bibitem[\protect\citeauthoryear{{Weinberg}, {Hernquist} \& {Katz}}{{Weinberg}
  et~al.}{2002}]{WeinbergHernquistKatz02}
{Weinberg} D.~H.,  {Hernquist} L.,    {Katz} N.,  2002, \apj, 571, 15

\end{thebibliography}

\end{document}